\documentclass[aps,prx,reprint,superscriptaddress,nofootinbib]{revtex4-2}

\usepackage{amsmath}

\usepackage{mathrsfs}

\usepackage{graphicx}

\usepackage[inline, shortlabels]{enumitem}

\usepackage{amssymb}

\usepackage{textgreek}

\usepackage{bm}

\usepackage{mhchem}

\usepackage{blindtext}

\usepackage{todonotes}

\usepackage{siunitx}
\DeclareSIUnit{\angstrom}{\textup{\AA}}

\usepackage{xspace}

\usepackage{import}

\usepackage{multirow}

\usepackage{setspace}

\usepackage[caption=false]{subfig}
\newcommand{\phantomsubfloat}[1]{
    {\captionsetup[subfigure]{labelformat=empty}
        \subfloat[][]{#1}
    }}

\usepackage{glossaries-extra}
\glsdisablehyper
\newcommand*\nestedglsentry[1]{\protect\ifglsused{#1}{\glsfmtshort{#1}}{\glsfmtlong{#1}}}
\setabbreviationstyle{long-short}
\setabbreviationstyle[ignored]{long-noshort}
\newabbreviation[category=ignored]{lro}{LRO}{long-range order}
\newabbreviation[category=ignored]{op}{OP}{order parameter}
\newabbreviation{cdw}{CDW}{charge density wave}
\newabbreviation{afm}{AFM}{antiferromagnet}
\newabbreviation{ife}{IFE}{inverse Faraday effect}
\newabbreviation{ptmb}{PTMB}{photo-thermal modulated birefringence}
\newabbreviation{xrd}{XRD}{X-ray diffraction}
\newabbreviation{shg}{SHG}{second harmonic generation}
\newabbreviation{trshg}{tr-SHG}{time-resolved \nestedglsentry{shg}}
\newabbreviation{rashg}{RA-SHG}{rotational-anisotropy \nestedglsentry{shg}}
\newabbreviation{paw}{PAW}{projected augmented wave method}
\newabbreviation{vasp}{VASP}{Vienna Ab-initio Software Package}
\newabbreviation{gga}{GGA}{generalized gradient approximation}
\newabbreviation{dft}{DFT}{density functional theory}
\newabbreviation{pdos}{PDOS}{partial density of states}
\newabbreviation{mae}{MAE}{magnetocrystalline anisotropy energy}
\newabbreviation{cmr}{CMR}{colossal magnetoresistance}
\newabbreviation{sc}{SC}{superconductivity}
\newabbreviation{tdgl}{TDGL}{time-dependent ginzburg landau}

\usepackage{hyperref}
\hypersetup{
    colorlinks=true,
    allcolors=blue,
}

\usepackage{cleveref}

\newcommand\fakesection[1]{
  \refstepcounter{section}
  \addcontentsline{toc}{section}{\protect\numberline{\thesection}#1}
  \sectionmark{#1}
}

\newcommand{\cmb}{\ce{CaMn2Bi2}\xspace}

\newcommand{\htpgmathmode}{\bar{3}m\xspace}
\newcommand{\htpg}{$\htpgmathmode$\xspace}
\newcommand{\degree}{^\circ}
\newcommand{\neel}{N\'{e}el\xspace}

\newcommand{\figs}[2]{Figs.~#1(#2)\xspace}
\newcommand{\onlinecref}[1]{Ref. \citenum{#1}\xspace}
\newcommand{\apx}{{\sim}\xspace}

\newcommand{\pt}{$\mathcal{PT}$\xspace}
\renewcommand{\sup}{Supplementary material}

\newcommand{\sfigonly}[1]{\sup, Fig~S#1\xspace}

\newcommand{\seqs}[2]{Eqs.~#1 and #2\xspace}

\newcommand{\PP}{P$_\mathrm{in}$--P$_\mathrm{out}$\xspace}
\newcommand{\PS}{P$_\mathrm{in}$--S$_\mathrm{out}$\xspace}
\newcommand{\SP}{S$_\mathrm{in}$--P$_\mathrm{out}$\xspace}
\renewcommand{\SS}{S$_\mathrm{in}$--S$_\mathrm{out}$\xspace}
\newcommand{\shortPP}{P_\mathrm{in}P_\mathrm{out}}
\newcommand{\shortPS}{P_\mathrm{in}S_\mathrm{out}}
\newcommand{\shortSP}{S_\mathrm{in}P_\mathrm{out}}
\newcommand{\shortSS}{S_\mathrm{in}S_\mathrm{out}}

\newcommand{\fulltempdep}{2}

\newcommand{\caplabel}{(\alph*)}
\newcommand{\capref}{(\alph*)}

\newcommand{\chione}{\chi^{(1)}}
\newcommand{\chitwo}{\chi^{(2)}}
\newcommand{\chithree}{\chi^{q}}
\newcommand{\chifour}{\chi^{m}}
\newcommand{\chithreeprime}{\chi^{q'}}
\newcommand{\chifourprime}{\chi^{m'}}
\newcommand{\threshold}{$\apx 200$ \si{\mu J \cdot cm^{-2}}\xspace}
\newcommand{\previousthreshold}{$\apx 10$ \si{mJ \cdot cm^{-2}}\xspace}
\newcommand{\supcref}[1]{\sup, \cref{#1}\xspace}
\newcommand{\supsupcref}[1]{\cref{#1}\xspace}

\crefname{figure}{Fig.}{Figs.}
\Crefname{figure}{Fig.}{Figs.}
\crefrangelabelformat{figure}{#3#1#4--#5#2#6}
\crefname{subfigure}{Fig.}{Figs.}
\Crefname{subfigure}{Fig.}{Figs.}
\crefrangelabelformat{subfigure}{#3#1#4--#5(\crefstripprefix{#1}{#2}#6}

\crefmultiformat{subfigure}{\edef\crefstripprefixinfo{#1}figs.~#2#1#3}{ and~#2\crefstripprefix{\crefstripprefixinfo}{#1}#3}{, #2\crefstripprefix{\crefstripprefixinfo}{#1}#3}{, and~#2\crefstripprefix{\crefstripprefixinfo}{#1}#3}

\makeatletter
\def\maketitle{
\@author@finish
\title@column\titleblock@produce
\suppressfloats[t]}
\makeatother

\begin{document}

\title{Light-induced reorientation transition in an antiferromagnetic semiconductor}

\author{Bryan T. Fichera}
\thanks{These authors contributed equally to this work}
\affiliation{Department of Physics, Massachusetts Institute of Technology, Cambridge, Massachusetts 02139}
\author{Baiqing Lv}
\thanks{These authors contributed equally to this work}
\affiliation{Department of Physics, Massachusetts Institute of Technology, Cambridge, Massachusetts 02139}
\affiliation{Department of Physics and Astronomy, Shanghai Jiao Tong University, Shanghai, 200240, China}
\author{Karna Morey}
\altaffiliation{Present address: Department of Physics, Stanford University, Stanford, CA 94305}
\affiliation{Department of Physics, Massachusetts Institute of Technology, Cambridge, Massachusetts 02139}
\author{Zongqi Shen}
\affiliation{Department of Physics, Massachusetts Institute of Technology, Cambridge, Massachusetts 02139}
\author{Changmin Lee}
\affiliation{Materials Science Division, Lawrence Berkeley National Laboratory, Berkeley, CA 94720}
\affiliation{Department of Physics, Hanyang University, Seoul 04763, Republic of Korea}
\author{Elizabeth Donoway}
\affiliation{Department of Physics, University of California at Berkeley, Berkeley, CA, USA}
\author{Alex Liebman-Pel\'{a}ez}
\affiliation{Department of Physics, University of California at Berkeley, Berkeley, CA, USA}
\author{Anshul Kogar}
\altaffiliation{Present address: Department of Physics and Astronomy, University of California, Los Angeles, CA, USA}
\affiliation{Department of Physics, Massachusetts Institute of Technology, Cambridge, Massachusetts 02139}
\author{Takashi Kurumaji}
\altaffiliation{Present address: Department of Advanced Materials Science, The University of Tokyo, Kashiwa 277-8561, Japan}
\affiliation{Department of Physics, Massachusetts Institute of Technology, Cambridge, Massachusetts 02139}
\author{Martin Rodriguez-Vega}
\affiliation{Department of Physics, Northeastern University, Boston, MA 02115, USA}
\affiliation{Department of Physics, The University of Texas at Austin, Austin, Texas 78712, USA}
\author{Rodrigo Humberto Aguilera del Toro}
\affiliation{Donostia International Physics Center (DIPC), 20018 Donostia-San Sebasti\'{a}n, Spain}
\affiliation{Centro de F'ısica de Materiales - Materials Physics Center (CFM-MPC), 20018 Donostia-San Sebasti\'{a}n, Spain}
\author{Mikel Arruabarrena}
\affiliation{Centro de F'ısica de Materiales - Materials Physics Center (CFM-MPC), 20018 Donostia-San Sebasti\'{a}n, Spain}
\author{Batyr Ilyas}
\affiliation{Department of Physics, Massachusetts Institute of Technology, Cambridge, Massachusetts 02139}
\author{Tianchuang Luo}
\affiliation{Department of Physics, Massachusetts Institute of Technology, Cambridge, Massachusetts 02139}
\author{Peter M\"{u}ller}
\affiliation{X-Ray Diffraction Facility, Department of Chemistry, Massachusetts Institute of Technology, Cambridge, Massachusetts 02139}
\author{Aritz Leonardo}
\affiliation{Donostia International Physics Center (DIPC), 20018 Donostia-San Sebasti\'{a}n, Spain}
\affiliation{EHU Quantum Center, University of the Basque Country UPV/EHU, 48940 Leioa, Spain}
\author{Andres Ayuela}
\affiliation{Donostia International Physics Center (DIPC), 20018 Donostia-San Sebasti\'{a}n, Spain}
\affiliation{Centro de F'ısica de Materiales - Materials Physics Center (CFM-MPC), 20018 Donostia-San Sebasti\'{a}n, Spain}
\author{Gregory A. Fiete}
\affiliation{Department of Physics, Massachusetts Institute of Technology, Cambridge, Massachusetts 02139}
\affiliation{Department of Physics, Northeastern University, Boston, MA 02115, USA}
\author{Joseph G. Checkelsky}
\affiliation{Department of Physics, Massachusetts Institute of Technology, Cambridge, Massachusetts 02139}
\author{Joseph Orenstein}
\affiliation{Materials Science Division, Lawrence Berkeley National Laboratory, Berkeley, CA 94720}
\affiliation{Department of Physics, University of California at Berkeley, Berkeley, CA, USA}
\author{Nuh Gedik}
\email{gedik@mit.edu}
\affiliation{Department of Physics, Massachusetts Institute of Technology, Cambridge, Massachusetts 02139}

\date{\today}

\begin{abstract}
Due to the lack of a net magnetic moment, antiferromagnets possess a unique robustness to external magnetic fields and are thus predicted to play an important role in future magnetic technologies.
However, this robustness also makes them quite difficult to control, and the development of novel methods to manipulate these systems with external stimuli is a fundamental goal of antiferromagnetic spintronics.
In this work, we report evidence for a metastable reorientation of the order parameter in an antiferromagnetic semiconductor triggered by an ultrafast quench of the equilibrium order via photoexcitation above the band gap.
The metastable state forms less than $10$ \si{ps} after the excitation pulse, and persists for longer than $150$ \si{ps} before decaying to the ground state via thermal fluctuations.
Importantly, this transition cannot be induced thermodynamically, and requires the system to be driven out of equilibrium.
Broadly speaking, this phenomenology is ultimately the result of large magnetoelastic coupling in combination with a relatively low symmetry of the magnetic ground state.
Since neither of these properties are particularly uncommon in magnetic materials, the observations presented here imply a generic path toward novel device technology enabled by ultrafast dynamics in antiferromagnets.
 \end{abstract}

\maketitle

\section{Introduction}

When the Hamiltonian of a system contains multiple interaction strengths of comparable magnitude, the corresponding free energy is often host to a diverse collection of metastable states just barely separated in energy from the true ground state.
This phenomenology results in an extreme sensitivity to external stimuli\cite{zhang_dynamics_2014,basov_electrodynamics_2011,dagotto_complexity_2005}, which can be exploited in an ultrafast way using light to drive transitions between these states\cite{basov_towards_2017,kogar_light-induced_2020,mitrano_possible_2016,fausti_light-induced_2011}.

One important application is in magnetic devices, where the electron spins form an ordered state in equilibrium and may thus be used to store information.
\Glspl{afm} have received special attention in this regard, as they possess zero net magnetic moment and are thus robust to stray magnetic fields from adjacent magnetic devices\cite{jungwirth_antiferromagnetic_2016}.
The dominance of exchange rather than anisotropy energies in the dynamics of spins which are antiferromagnetically ordered also leads to order-of-magnitude faster switching timescales compared to their ferromagnetic counterparts\cite{nemec_antiferromagnetic_2018}.

Importantly, the ground state of an \gls{afm} is typically not unique, with the number of degenerate states (corresponding generically to various rotations of the \gls{afm} \gls{op}) being determined by the number of symmetries which are broken at the magnetic ordering temperature.
In the context described above, this degeneracy invites the possibility of an ultrafast antiferromagnetic device which uses light to switch between these different states.
Moreover, if the symmetry of the magnetic phase is sufficiently low (relative to the parent phase), the number of such states can be quite high, and a multi-step magnetic memory might be constructed which operates via ultrafast reorientation transitions of the \gls{afm} \gls{op} between states that are not anti-parallel.

Broadly speaking, one can distinguish two different approaches for achieving such transitions in real materials.
In the first case, one imagines manipulating the \gls{afm} orientation ``coherently'' by, for example, resonantly driving an infrared-active phonon mode to large amplitude, although this usually requires large electric fields (as the relevant light-matter interactions are usually highly nonlinear\cite{afanasiev_ultrafast_2021}) which are difficult to obtain in the frequency range associated with phonons in crystals.
An alternative mechanism---which has been shown to occur abundantly in \gls{sc}\cite{fausti_light-induced_2011} and \gls{cdw}\cite{kogar_light-induced_2020} systems---is to \textit{quench} the equilibrium order by exciting carriers above the band gap, and then exploit the subsequent relaxation dynamics which may, in the right system, show a preference for one state over the other\cite{sun_transient_2020}.
In contrast to the case of resonant phonon driving, this mechanism involves primarily electronic excitations and therefore requires relatively little energy if the photon frequency is above the band gap.
However, despite successful demonstrations of this approach in nonmagnetic systems\cite{fausti_light-induced_2011,kogar_light-induced_2020}, evidence for this mechanism occurring in \glspl{afm} does not currently exist, and whether the phenomenology described here actually occurs in real magnetic materials remains a fundamental open question.

In this work, we report a light-induced phase transition between nearly degenerate \gls{afm} states in the antiferromagnetic semiconductor \cmb \cite{gibson_magnetic_2015} triggered by an ultrafast quench of the equilibrium \gls{op} with a femtosecond light pulse.
Using \gls{trshg}, a pump-probe technique sensitive to the magnitude and direction of the \gls{afm} order parameter, we find that above-gap optical excitation indeed causes the system to reorient to a new, metastable orientation of the \gls{afm} \gls{op}.
The transition is fast (occurring less than $10$ \si{ps} after photoexcitation), requires low pump fluence (\threshold vs. \previousthreshold\cite{afanasiev_ultrafast_2021}), and cannot be induced thermodynamically.
These findings suggest that the same post-quench, metastable dynamics occuring in nonmagnetic systems\cite{kogar_light-induced_2020,fausti_light-induced_2011} applies to \glspl{afm} as well, opening the door to potential opto-spintronic device architectures exploiting nonequilibrium phase transitions between different nearly degenerate states. 
 
\section{Results}

\subsection{Equilibrium}

\begin{figure}
\centering{
\phantomsubfloat{\label{fig0:a}}
\phantomsubfloat{\label{fig0:b}}
\phantomsubfloat{\label{fig0:c}}
\phantomsubfloat{\label{fig0:d}}
}
\includegraphics{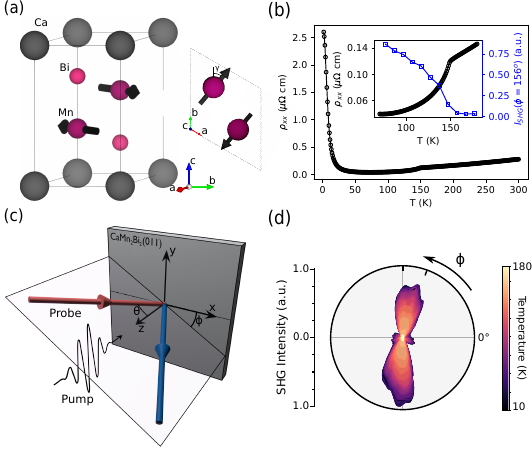}
\captionsetup{singlelinecheck=off}
\caption[]{
\label{fig0}
\begin{enumerate*}[label=\caplabel, ref=\capref]
\item Magnetic unit cell of \cmb.
The angle $\gamma$ is such that the \ce{Mn} spin does not lie precisely along any particular crystallographic axis.
\item Resistivity as a function of temperature for a representative sample.
Inset shows an enlarged view of the resistivity near $T_c$ plotted against the SHG intensity integrated across the region indicated in \ref{fig0:d}.
\item Schematic of the \glsfmtshort{rashg} setup.
The crystallographic $a$ axis lies along the dashed line in the figure (see \supcref{sec:orientation}).
\item Temperature dependence of the \glsfmtshort{rashg} intensity in the \PP polarization channel from the $(011)$ surface of \cmb.
Other polarization channels are shown in the \sfigonly{\fulltempdep}.
The area between the two ticks indicates the region integrated to produce the inset of \ref{fig0:b}.
\end{enumerate*}
}
\end{figure}
 
\begin{figure*}
\centering{
\phantomsubfloat{\label{fig1:a}}
\phantomsubfloat{\label{fig1:b}}
\phantomsubfloat{\label{fig1:c}}
\phantomsubfloat{\label{fig1:d}}
\phantomsubfloat{\label{fig1:e}}
\phantomsubfloat{\label{fig1:f}}
\phantomsubfloat{\label{fig1:g}}
}
\includegraphics{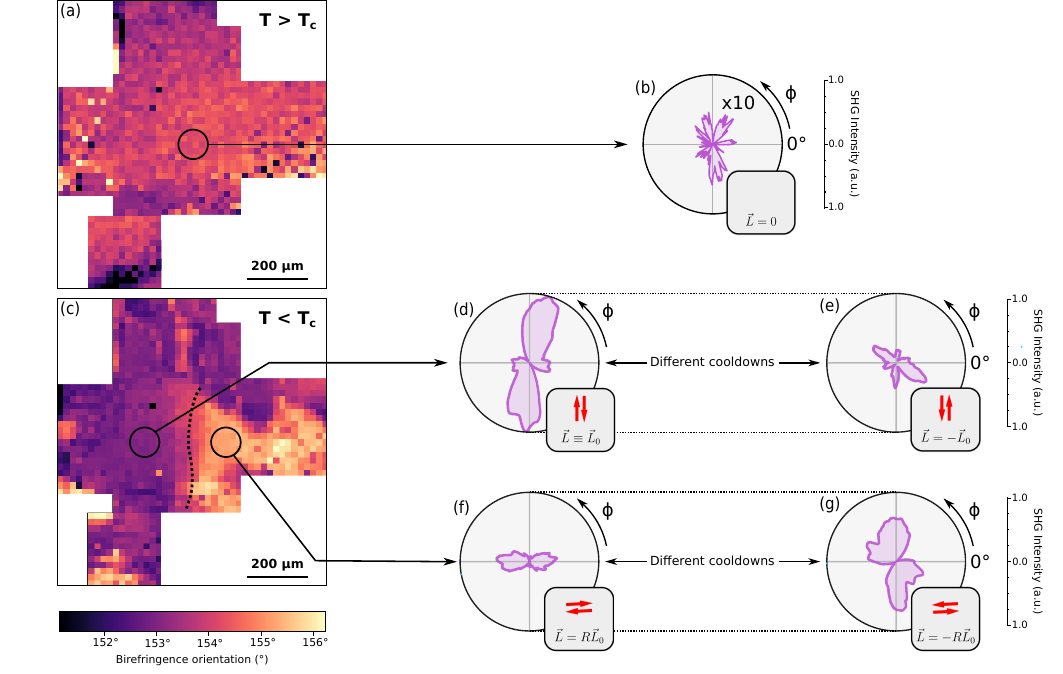}
\captionsetup{singlelinecheck=off}
\caption[]{
\label{fig1}
\begin{enumerate*}[label=\caplabel, ref=\capref]
\item Birefringence orientation map for $T=200~\si{K}>T_c$.
\item \glsfmtshort{rashg} intensity in the \PP polarization channel for the region indicated in \ref{fig1:a} at $157$ \si{K}.
Other polarization channels are shown in the \supcref{fullequilibrium}.
\item Birefringence orientation map for $T=152.5~\si{K}<T_c$.
\item {} \item \glsfmtshort{rashg} intensity at $8$ \si{K} in the \PP polarization channel for consecutive cooldowns for the region indicated in \ref{fig1:c}.
The system randomly chooses the \glsfmtshort{rashg} pattern in \ref{fig1:d} or \ref{fig1:e} on each cooldown.
\item {} \item \glsfmtshort{rashg} intensity at $8$ \si{K} in the \PP polarization channel for consecutive cooldowns for the region indicated in \ref{fig1:c}.
The system randomly chooses the \glsfmtshort{rashg} pattern in \ref{fig1:f} or \ref{fig1:g} on each cooldown.
Other polarization channels are shown in the \supcref{fullequilibrium}.
The insets beneath the \glsfmtshort{rashg} plots depict the relationships between the \neel vectors $\vec{L}$ associated with each domain, as determined from the \glsfmtshort{rashg} and \glsfmtshort{ptmb} data: \glsfmtshort{rashg} patterns on the same spot but different cooldowns correspond to \neel vectors which are related by an overall sign ($\vec{L}=\pm\vec{L_0}$ and $\vec{L}=\pm R\vec{L_0}$), whereas those on different spots are related by an element $R$ (not identity or inversion) of the high-symmetry point group $D_{3d}$.
Neither the value of $\vec{L}_0$ nor the operation $R$ are explicitly determined by these measurements.
Note that the \glsfmtshort{rashg} experiments are performed normal to the (011) plane (see \cref{fig0}), whereas the vectors in the insets are drawn in the (001) plane.
\end{enumerate*}
}
\end{figure*}
 
Crystallographically, \cmb consists of a puckered-honeycomb tiling of \ce{Mn} atoms with a high-temperature crystallographic point group of \htpg (see \cref{fig0:a}).
The electronic gap is on the order of $4$ \si{meV} (see \cref{fig0:b} and \supcref{transport:a}) and is thought to be due to a delicate combination of correlations and hybridization between relatively localized \ce{Mn} $3d$ states and dispersive \ce{Bi} $6p$ / \ce{Mn} $4s$ hybrid bands\cite{gibson_magnetic_2015, piva_putative_2019, lane_competition_2019}.
N\'{e}el-type antiferromagnetic \gls{lro} with spins lying in the $ab$-plane develops in this material at $T_c=\apx150$ \si{K}, accompanied by a kink in the electrical resistivity (\cref{fig0:b}) and magnetization (\supcref{transport:b}) and the appearance of a new peak in the powder neutron diffraction\cite{gibson_magnetic_2015}.
Importantly, the powder neutron refinement indicates that the ordered moments are slightly misaligned with the $a$- and $b$-axes of the high-symmetry phase (see \cref{fig0:a}), so that the low-temperature magnetic point group is $\bar{1}'$; i.e., the only symmetries that are preserved in the low temperature phase are the identity operator and the product of inversion and time-reversal\cite{gibson_magnetic_2015}.
The remarkably low symmetry of this magnetic ground state is likely due to highly frustrated magnetic interactions, as suggested by the proximity of isostructural \ce{CaMn2Sb2} to a mean-field magnetic tricritical point\cite{mazin_camn2sb2_2013,mcnally_camn2sb2_2015}.
In contrast, the lattice is thought to remain fully symmetric at all temperatures, so that the symmetry-breaking at $T_c$ is solely due to the \gls{afm} order (this is verified with single-crystal \gls{xrd}, see \supcref{xrd}).

As spin-orbit coupling is expected to be strong in this material, the breaking of inversion symmetry by the magnetic order should result in a strong \gls{shg} signal below $T_c$, despite the fact that the lattice remains centrosymmetric\cite{pan_optical_1989,seyler_spin-orbit-enhanced_2020}.
We probe this \gls{shg} signal using a \gls{rashg} apparatus (\cref{fig0:c}) which measures the reflected second harmonic intensity as the plane of incidence (and the incoming and outgoing polarization vectors, which may be P- or S-polarized relative to the plane of incidence) is rotated about the sample normal\cite{fichera_second_2020, harter_high-speed_2015, torchinsky_low_2014}.
\Cref{fig0:b,fig0:d} depict the \gls{shg} signal from the $(011)$ surface of \cmb as a function of temperature across $T_c$, demonstrating that \gls{rashg} is indeed sensitive to the magnetic order in this material.
 
Frequently, \gls{afm} materials may host intricate domain configurations, the details of which greatly impact the relevant magnetic functionalities\cite{reimers_defect-driven_2022, weber_magnetostrictive_2003}.
Because the low-temperature magnetic point group of \cmb breaks multiple symmetries of the high-temperature point group, we indeed expect that the low-temperature magnetic ground state should involve some number of energetically degenerate domains.
In \Cref{fig1}, we characterize these domains using a combination of \gls{rashg} and a spatially-resolved optical polarimetry technique known as \gls{ptmb}, which is a sensitive technique for measuring small changes in the anisotropic index of refraction (and consequently, the direction of the \gls{afm} \gls{op})\cite{little_three-state_2020, lee_observation_2022}.

\cref{fig1:a} shows the \gls{ptmb} signal from the $(011)$ surface of \cmb above $T_c$ (see \supcref{sec:hightempptmb} and \supcref{birefringence}).
No contrast is found within a $\apx500\si{\mu m} \times 500\si{\mu m}$ area on the sample, demonstrating that the thermally modulated index of refraction is spatially uniform at high temperature.
As the temperature is lowered below $T_c$ (\cref{fig1:c}), a pronounced contrast appears in the \gls{ptmb} map which indicates the presence of two separate \gls{afm} domains with different \gls{op} directions.

While \gls{ptmb} is nominally sensitive to the direction of the \gls{afm} \gls{op}, it cannot differentiate between $180\degree$ domains as it couples only to the linear index of refraction of the material.
Therefore, it is not clear from \gls{ptmb} alone whether the domains in \cref{fig1:c} are truly homogeneous or may contain a mixture of $180\degree$ domains.
In contrast, nonlinear spectroscopies like \gls{shg} can differentiate $180\degree$ \gls{afm} domains due to an interference between the \gls{shg} sources (e.g. electric dipole, magnetic dipole, and electric quadrupole, see \supcref{interference}) which transform differently under time-reversal symmetry\cite{fiebig_second-harmonic_2005, fiebig_second_1994, fiebig_second_2001, fiebig_domain_1995}.
This effect can be especially pronounced in the presence of magnetostriction\cite{fiebig_second_2001}.
We thus proceed to investigate the presence of $180\degree$ domains in this material using \gls{rashg}.

\crefrange{fig1:d}{fig1:g} depict the \gls{rashg} results at $8$ \si{K} in the two regions identified in \cref{fig1:c}.
The same crystal piece was used for both measurements.
As with \gls{ptmb}, we do find that the \gls{shg} domains are large and apparently homogenous (on a scale of hundreds of \si{\mu m}, see \supcref{domain_homogeneity}).
Surprisingly, however, in different cooldowns we observe that the \gls{rashg} pattern in the leftmost region of \cref{fig1:c} randomly takes the form of either \cref{fig1:d} or \cref{fig1:e}, and in the rightmost region, it takes the form of either \cref{fig1:f} or \cref{fig1:g}.
In contrast, the \gls{ptmb} does not change upon thermal cycling (see \supcref{thermalcycles}).

Since the four domains in \cref{fig1:d,fig1:e,fig1:f,fig1:g} have different \gls{rashg} patterns, they correspond to four different \gls{afm} domains with different values of the \neel vector (defined as the difference in the magnetic moment on the two \ce{Mn} sublattices, see \cref{fig0:a}).
While the \gls{rashg} refinement (see \supcref{sec:fitting}) does not explicitly determine the value of the \neel vector in each of these domains, the fact that the \gls{rashg} pattern changes on thermal cycling while the \gls{ptmb} signal does not indicates that each pair of domains in \cref{fig1} consists of two domains with \ang{180}-opposite \neel vectors.
Furthermore, there is a relative angle between \crefrange{fig1:d}{fig1:e} and \crefrange{fig1:f}{fig1:g}, since they have different \gls{ptmb} signals.\footnote{Note that, in general, we do not expect the direction of the \neel vector to be qualitatively apparent from the RA-SHG pattern, since the (001) easy plane of the \glsfmtshort{afm} order is not the plane which is normal to the optical axis (which is (011), the natural cleavage plane of \cmb, in our measurement geometry; see \cref{fig0}).}
An example of a domain assignment satisfying these criteria is depicted schematically in the insets to \crefrange{fig1:d}{fig1:g}, where the \gls{afm} \gls{op} is represented by two red arrows.
Importantly, within a single location, the two opposite orientations are accessible via thermal cycling; hence they are energetically degenerate.
We have verified that the relationship between domains described here is consistent with the \gls{shg} susceptibility tensors\cite{boyd} extracted by fitting the data in \cref{fig1} (see \supcref{sec:fitting}).
In addition, our data is not consistent with an alternative interpretation involving the interference of two independent order parameters (see \supcref{sec:alternativeinterpretation}).
 
\subsection{Nonequilibrium}

\begin{figure*}
\centering{
\phantomsubfloat{\label{fig2:a}}
\phantomsubfloat{\label{fig2:b}}
\phantomsubfloat{\label{fig2:c}}
\phantomsubfloat{\label{fig2:d}}
}
\includegraphics{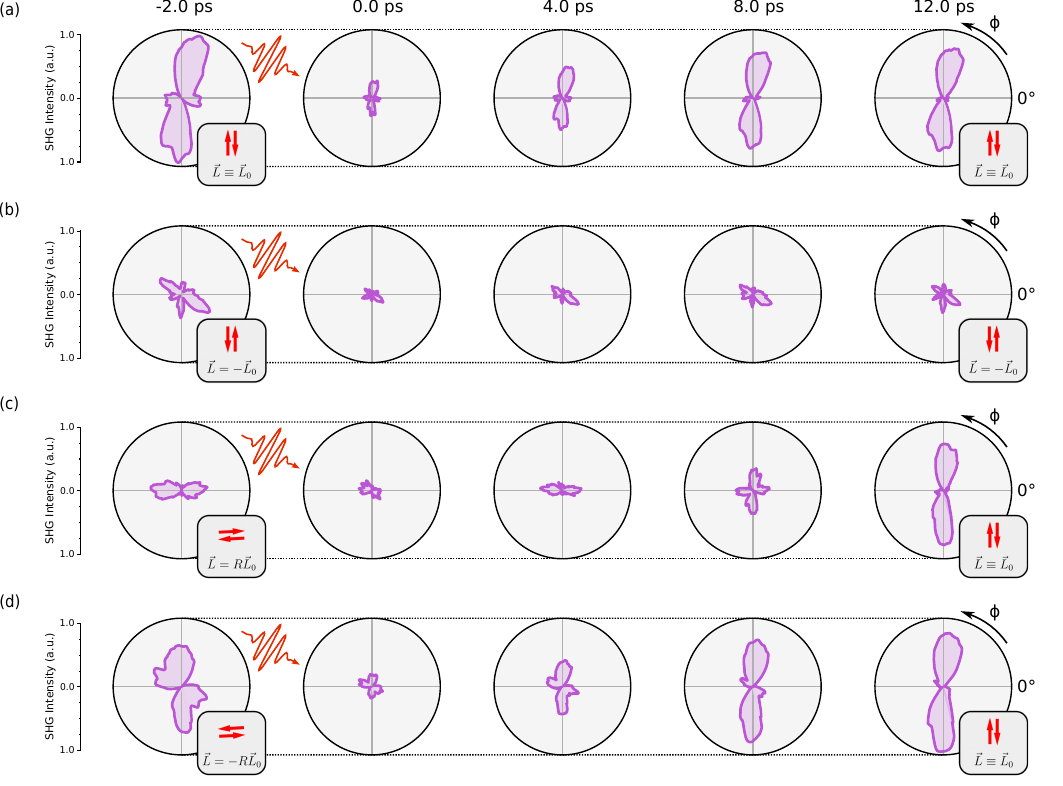}
\captionsetup{singlelinecheck=off}
\caption[]{
\label{fig2}
\begin{enumerate*}[label=\caplabel, ref=\capref]
\item[] \glsfmtshort{rashg} intensity at $8$ \si{K} in the \PP polarization channel as a function of time for the \glsfmtshort{afm} \glsfmtshort{op} identified in \item \cref{fig1:d}, \item \cref{fig1:e}, \item \cref{fig1:f}, and \item \cref{fig1:g}.
Other polarization channels are shown in the \supcref{fullnonequilibrium}.
The pump fluence is set to $\apx 600$ \si{\mu J \cdot cm^{-2}}.
The insets beneath the \glsfmtshort{rashg} plots depict schematically the value of the \neel vector \qty{2}{ps} before and \qty{12}{ps} after photoexcitation, as in \cref{fig1}.
\end{enumerate*}
}
\end{figure*}
 
Having demonstrated that \cmb hosts an elaborate free energy surface with multiple degenerate or nearly degenerate ground states, we now turn to the question of whether light can be used to manipulate or possibly switch between these states. 
To answer this question, we pump each of the \gls{op} directions identified in \cref{fig1} at $8$ \si{K} with a $\apx 100$ \si{fs} normally incident near-infrared light pulse ($\hbar \omega \approx 1$ \si{eV}) and probe the \gls{afm} \gls{op} with \gls{rashg} using a subsequent probe pulse.
By varying the delay $\Delta t$ between the two pulses, we create a series of snapshots of the \gls{afm} \gls{op} at different times following excitation.
The results are shown in \cref{fig2}, where the four rows correspond to the four \gls{op} directions identified in \cref{fig1}, and the horizontal axis is the time delay between the pump and probe pulses.
The first two rows (\crefrange{fig2:a}{fig2:b}) correspond to the leftmost domain in \cref{fig1:c}, for which it is found that, after pumping, the strength of the \gls{rashg} signal is quickly extinguished.
After a delay of around $\apx 8-10$ \si{ps}, the \gls{afm} order recovers so that the shape of the \gls{rashg} pattern is similar to before zero delay up to small changes in the relative peak heights.
The \gls{op} magnitude and direction inferred by this data is depicted schematically in the insets to \crefrange{fig2:a}{fig2:b}.

The striking observation is in the latter two rows (\crefrange{fig2:c}{fig2:d}), which depict the time-resolved \gls{rashg} results in the rightmost domain of \cref{fig1:c}.
In this domain, the \gls{rashg} signal again quickly decreases, and the shape does not initially change.
However, roughly $\apx 6-8$ \si{ps} after excitation, the \gls{rashg} pattern abruptly changes shape, so that the pattern after $\apx 10$ \si{ps} is equivalent to \cref{fig2:a}.
Together with the findings of \cref{fig1}, these results suggest that the final state in \crefrange{fig2:c}{fig2:d} is one in which the \gls{afm} \gls{op} direction has indeed been reoriented relative to equilibrium.
Remarkably, one must use light to reach this metastable state as it is not present in thermal equilibrium at this spot.

We performed the above measurements out to $\Delta t \gtrsim 150$ \si{ps} (see \supcref{CtoBlong}) and found that the reoriented state persists at least this long before relaxing back to equilibrium in time for the next pair of pulses to arrive (roughly $200$ \si{\mu s} later).
As the pump fluence is decreased (see \supcref{fluencedep}), the effect of the excitation is diminished until, at low fluences ($\apx 100$ \si{\mu J \cdot cm^{-2}}), the final state is equivalent to the initial state, and only small changes to the \gls{shg} (associated with dynamics that do not launch the system into the metastable state) are visible near zero delay.
Interestingly, which of the two opposite directions (\cref{fig1:d} or \cref{fig1:e}) the magnetic order favors after excitation is consistent from pulse to pulse, but not from one sample to another (see \supcref{CtoBshort,CtoBlong}), suggesting that, below the magnetic ordering temperature, interactions between neighboring domains break the degeneracy between these states.
Finally, we note that no aspect of these observations changes with the polarization of the pump pulse (see \supcref{polarization}).
 
\section{Discussion}

\begin{figure*}
\centering{
\phantomsubfloat{\label{quenchtimes}}
\phantomsubfloat{\label{fig3:a}}
\phantomsubfloat{\label{fig3:b}}
\phantomsubfloat{\label{fig3:c}}
\phantomsubfloat{\label{fig3:d}}
\phantomsubfloat{\label{fig3:e}}
}
\includegraphics{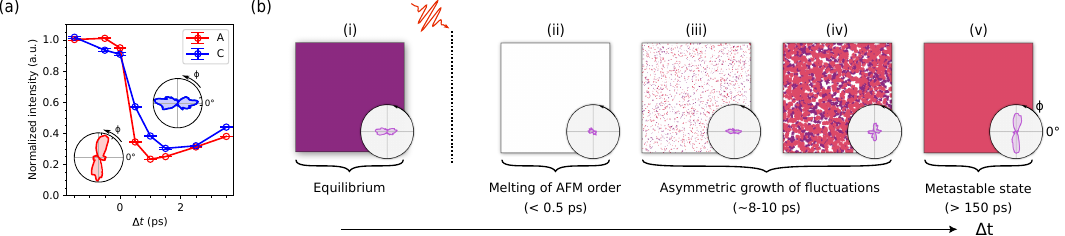}
\captionsetup{singlelinecheck=off}
\caption[]{
\label{fig3}
\begin{enumerate*}[label=\caplabel, ref=\capref]
\item Integrated \glsfmtshort{shg} intensity as a function of time corresponding to \cref{fig2:a} (A, red) and \cref{fig2:c} (C, blue).
Error bars were calculated as described in \supcref{sec:errorbars}.
Insets show the equilibrium \glsfmtshort{shg} patterns in the \PP polarization combination for the two domains, with the integration region indicated.
The $y$-axis normalization for either domain is defined such that the average signal before zero delay is equal to $1.0$.
\item Conceptual illustration of the dynamics following laser excitation as described in the text.
Ultrafast melting of the \glsfmtshort{afm} order (i-ii) is followed by the rapid growth of fluctuations (iii-iv), with the final state (v) determined by the relative growth rate of different orders.
Insets show the \glsfmtshort{rashg} data of \cref{fig2:c} corresponding roughly to each of these steps.
The fluctuation growth step (iii-iv) may occur with or without a background of coherent oscillations (see \supcref{sec:fluctuations,sec:nonequilibriumtheory}).
\end{enumerate*}
}
\end{figure*}
 
Taken together, these considerations indeed point to a \textit{bona fide} ultrafast phase transition in \cmb induced by a femtosecond light pulse.
We now seek to understand, at least qualitatively, the microscopic nature of this transition.
Before discussing the time-resolved data of \cref{fig2}, we must first understand the equilibrium phenomenology implied by \cref{fig1}.
Group theory (see \supcref{sec:nonequilibriumtheory}) suggests that there should be twelve energetically degenerate \gls{op} directions (see \supcref{fig:grouptheory}), independent of location on the sample.
However, only two directions are observed per measurement location in our experiment.
This observation is explained by recognizing that local strain fields in the material are expected to pole the \gls{lro} (see \supcref{sec:equilibriumtheory}), so that certain \gls{op} directions are energetically favored relative to others at a given spot\cite{higo_perpendicular_2022,ni_imaging_2021,zhu_strain-modulated_2014,koziol-rachwal_control_2020}.
Since these strain fields may vary from one location to another (the details of which depend subtly on the local growth conditions), different locations on the sample would then show different sets of \gls{rashg} patterns, in agreement with the findings of \cref{fig1}.
This picture also explains the observation that opposite \gls{afm} configurations are energetically degenerate, as strain itself is inherently symmetric under $180^\circ$ rotations.

Turning now to the nonequilibrium phenomenology (\cref{fig2,fig3}), we begin by noting that the sudden suppression of the \gls{shg} intensity at early times indicates that the primary effect of the pump pulse is simply to quench the \gls{afm} \gls{op} (\cref{fig3:a,fig3:b}).
According to recent theoretical work based on the \gls{tdgl} equation, \cite{sun_transient_2020, dolgirev_self-similar_2020}, the dynamics following such a quench are governed not by the global minimum of the free energy, but rather by the exponential amplification of spatial \gls{op} fluctuations (\cref{fig3:c,fig3:d}), which may occur differently for different orders (\supcref{sec:fluctuations}).
The final state of the transition (\cref{fig3:e}) is then determined by which state supports the fastest-growing fluctuations.
In the context of light-induced \gls{sc}\cite{fausti_light-induced_2011,cremin_photoenhanced_2019}, for example, the dominant order in equilibrium is a \gls{cdw}, whose fluctuations naturally involve motion of the massive nuclei and are thus much slower than fluctuations of the \gls{sc} \gls{op}\cite{smallwood_tracking_2012,sun_transient_2020}.
Note that, in the general case, the final state is determined not just by the relaxation rates, but also by their start times; that is, if one order is \textit{quenched} faster than another, fluctuations of that order are afforded a longer period to grow and may thus dominate at long times.

In our system, the relatively slow ($\apx 8-10$ \si{ps}, see \cref{fig2}) recovery timescale indicates that the dynamics of the \gls{afm} \gls{op}, like the \gls{cdw} order of \onlinecref{fausti_light-induced_2011}, also likely involve motion of the atomic nuclei (i.e. magnetoelastic coupling).
However, in contrast to the \gls{cdw}-\gls{sc} competition referred to above, the dominant and subdominant orders in our system correspond not to completely different orders, but rather to different orientations of the same order.
One would thus naively expect the corresponding fluctuations to exhibit more or less the same dynamics.
However, as described above, this naive symmetry of the material is in fact broken at all temperatures by the internal strain; it is therefore appropriate to think of the two non-parallel \gls{afm} orientations as truly distinct orders, and the \gls{tdgl} analysis described above thus applies rigorously to \cmb.
The situation is similar to the case of \ce{LaTe3}, where the equilibrium and nonequilibrium orders are both \glspl{cdw} which are split by a small lattice anisotropy\citep{kogar_light-induced_2020}.
Importantly, it is noted in \onlinecref{sun_transient_2020} that small variations in the density of different fluctuations are exponentially amplified as the system relaxes, so that only a modest difference in the relevant dynamics is sufficient to favor the metastable state as the system recovers.
Such asymmetric amplification of non-parallel \gls{afm} orientations thus likely forms the basis for the reorientation transition in \cmb.

Additional evidence supporting this general picture may be found by examining the time dependence of the \gls{shg} signal at short times (\cref{quenchtimes}), before the \gls{op} switches to the metastable state.
In particular, on examining the initial decay rates it can be seen that the \gls{shg} signal corresponding to \cref{fig2:a} reaches $30\%$ of its original value just $0.5$ \si{ps} after zero delay; the equivalent drop in \gls{shg} intensity for \cref{fig2:c} takes $> 0.5$ \si{ps} longer.
That is, the time required to quench the final state is approximately half the time needed to quench the initial state.
As a result, long-wavelength fluctuations of the subdominant, metastable order experience a longer period of exponential amplification compared to the equilibrium order, which may explain the dominance of the metastable state at long times (see \onlinecref{sun_transient_2020}, Sec. V).
The metastability of the final state is provided by the fact that the long-term relaxation dynamics are presumably dependent on the thermal nucleation and subsequent motion of \gls{afm} domain walls, which can be quite slow\cite{parkin_memory_2015,kim_real-space_2022}.
Finally, we remark that while the quench dynamics described here can explain all features of our data, they also likely occur alongside a background of coherent excitations (e.g. coherent phonons, which are observable in time-resolved reflectivity, see \supcref{pp}) which can couple to the order parameter via the lattice strain (see \supcref{sec:nonequilibriumtheory}).
While no evidence of these excitations is apparent in our \gls{shg} data, further studies will be needed to determine the extent to which they might still be important.
 
\section{Conclusion}

To summarize, we have presented evidence of a magnetic reorientation transition in the antiferromagnetic semiconductor \cmb induced by above-gap optical excitation.
Looking forward, we note two ingredients that seem to be most important for the transition demonstrated in this work: low symmetry, so that there are multiple energetic minima in the magnetic phase, and, presumably, some degree of magnetoelastic coupling, so that the different minima are not precisely degenerate in thermal equilibrium.
Neither of these properties are unique to \cmb, and analogous materials may exist with similar light-induced phenomenology. 
The apparently pronounced coupling between light and magnetic order in this material also invites the fascinating possibility that related phenomenology could also be accessed via additional tuning parameters like strain or current.
This is a key prediction of so-called \pt-symmetric systems, in which inversion and time-reversal symmetries are independently broken but the product of the two is preserved\cite{watanabe_group-theoretical_2018, watanabe_nonlinear_2020, watanabe_symmetry_2018}.
Further investigation of these and related compounds could have far-reaching impacts for future opto-spintronic technology.
 
\section{Methods}

\subsection{Material growth}

Single crystals of \cmb were grown by a \ce{Bi} flux method following \onlinecref{gibson_magnetic_2015}.
The starting materials were \ce{Ca} ingot, \ce{Mn} powder, and \ce{Bi} powder.
They were loaded into an alumina crucible in the molar ratio of $\ce{Ca}:\ce{Mn}:\ce{Bi} = 1:2:10$, which was sealed in an evacuated quartz tube.
The ampoule was heated to $1000^\circ$\si{C} and kept at this temperature for $11$ hours before being slowly cooled to $400^\circ$\si{C} in $10$ days.
The excess flux was removed by a centrifuge at this temperature.
The single phase nature of the crystals was checked by powder \gls{xrd} and the orientation was checked by a single-crystal x-ray diffractometer.

\subsection{\gls{trshg} measurements}

The \gls{trshg} measurements were performed using a fast-rotating optical grating method---where the plane of incidence is rotated about the normal axis so as to access all elements of the \gls{shg} susceptibility tensor---which has been described in detail elsewhere\cite{fichera_second_2020, harter_high-speed_2015, torchinsky_low_2014}.
Ultrashort ($\apx 100$ \si{fs}) probe pulses were sourced from a regeneratively amplified \ce{Ti}:Sapphire laser operating at a repetition rate of $5$ \si{kHz}.
A portion of the \ce{Ti}:Sapphire output was directed to an optical parametric amplifier, producing $\apx 1$ \si{eV} pump pulses which were delayed relative to the probe pulses with an optical delay line.
The incident probe fluence was $\apx 250$ \si{\mu J \cdot cm^{-2}}, with the probe spot diameter being $\apx 100$ \si{\mu m} at the sample surface, and the pump pulses (with varying fluence) were focused to a spot roughly $\apx 400$ \si{\mu m} in diameter.
Note that all of the \gls{shg} intensity is generated instantaneously when the probe interacts with the sample, so any subsequent dynamics induced by the probe pulse itself are not measured by our detector.
We verified that the pump-induced static heating at these fluences was insignificant by measuring the temperature dependence of the \gls{shg} signal with and without the pump, and noting that the transition temperature is not meaningfully different (see \cref{pump-on-tempdep}).
The (011) plane was found to be the only natural cleavage plane of the sample.
The pump was normally incident to the sample surface (to facilitate blocking the reflected pump scatter), and the probe was incident with an angle of $\apx 10^\circ$.
The reflected \gls{shg} signal was collected with a photomultiplier tube, filtered with a lock-in amplifier operating at the repetition rate of the regenerative amplifier, and correlated with the rotation angle of the grating with an optical rotary encoder and homebuilt oscilloscope.
To eliminate potential artifacts due to low-frequency fluctuations in the laser intensity, multiple random sweeps of the delay stage were averaged together for each dataset, and the polarizers were controlled automatically at each delay using custom polarization rotators described in \onlinecref{morey_in_prep}.

\subsection{\gls{ptmb} measurements}

The polarization-dependent optical birefringence (linear dichroism) measurements were conducted using the experimental setup detailed in \onlinecref{little_three-state_2020}.
A $633$ \si{nm} probe laser beam was focused onto a $5$ \si{\mu m} spot at surface of the sample, while a $780$ \si{nm} pump/heating beam was similarly focused onto the same location.
To improve the signal-to-noise ratio and eliminate polarization artifacts originating from the experimental setup, the sample temperature was modulated at $2$ \si{kHz} by passing the pump beam through an optical chopper.
The linear dichroism signal was subsequently measured using a standard balanced photodetection scheme and a lock-in amplifier.
Position-dependent measurements of optical birefringence were obtained by scanning the sample position with a piezoelectric scanner.
The dependence of the polarization rotation on the probe polarization angle was fit to a sinusoidal function at each spot, and the phase shift parameter extracted from each fit was plotted in \cref{fig1:a,fig1:c} (see \supcref{birefringence}).
Note that the birefringence signal involves contributions from both the lattice and the magnetism, so that the angle extracted from this fit should not be identified with the angle of the \gls{afm} \gls{op}.

\subsection{\gls{xrd} measurements}

\Gls{xrd} data (\supcref{xrd}) were collected at $100$ \si{K} and $180$ \si{K} on a Bruker-AXS X8 Kappa Duo diffractometer with I\textmu S micro-sources, coupled to a Photon 3 CPAD detector using \si{Mo} K\textalpha~radiation ($\lambda = 0.71073$ \si{\angstrom}), performing $\phi$-and $\omega$-scans. 
Reconstructed precession images of the $(0kl)$, $(h0l)$ and $(hk0)$ planes were calculated directly from the diffraction images using algorithms included in the APEX3\cite{apex} software.  

\subsection{Data availability}

Data supporting the figures within this paper and other findings of this study are available from the corresponding author upon request.
 
\begin{acknowledgments}
The authors thank Clifford Allington, Martin Eckstein, Shiang Fang, Feng Hao, David Hsieh, Honglie Ning, Jia Xu, and Guangua Zhang for several helpful discussions.
The authors also acknowledge the MIT SuperCloud and Lincoln Laboratory Supercomputing Center for providing HPC resources that have contributed to the research results reported within this paper.
B.F., B.L., K.M., Z.S., B.I., T.L., and N.G. acknowledge support from the US Department of Energy, BES DMSE (data taking and analysis), and Gordon and Betty Moore Foundation’s EPiQS Initiative grant GBMF9459 (instrumentation).
C.L., E.D., A.L.-P., and J.O. acknowledge support from the Quantum Materials program under the Director, Office of Science, Office of Basic Energy Sciences, Materials Sciences and Engineering Division, of the US Department of Energy, Contract No. DE-AC02-05CH11231.
G.A.F. gratefully acknowledges support from the NSF through the Center for Dynamics and Control of Materials: an NSF MRSEC under DMR-1720595, NSF DMR2114825, and the Alexander von Humboldt Foundation.
R.H.A.d.T., M.A.,A.L., and A.A. acknowledge support from the Spanish Ministry of Science and Innovation through grants PID2019-105488GB-I00, TED2021-132074B-C32 and PID2022-PID2022-139230NB-I00, the European Commission from the MIRACLE (ID 964450), and the Basque Government through Project No. IT-1569-22.
 \end{acknowledgments}

\clearpage

\fakesection{sup}

\renewcommand{\thepage}{S\arabic{page}} 
\renewcommand{\thesection}{S\arabic{section}}  
\renewcommand{\thesubsection}{S\arabic{subsection}}  
\renewcommand{\thesubsubsection}{\arabic{subsubsection}}  
\renewcommand{\thefigure}{S\arabic{figure}} 
\makeatletter
\renewcommand{\p@subsection}{}
\renewcommand{\p@subsubsection}{\thesubsection.}
\makeatother
\setcounter{figure}{0}
\setcounter{section}{0}
\setcounter{subsection}{0}
\setcounter{page}{1}

\title{Supplementary material for ``Light-induced reorientation transition in an antiferromagnetic semiconductor''}

\author{Bryan T. Fichera}
\thanks{These authors contributed equally to this work}
\affiliation{Department of Physics, Massachusetts Institute of Technology, Cambridge, Massachusetts 02139}
\author{Baiqing Lv}
\thanks{These authors contributed equally to this work}
\affiliation{Department of Physics, Massachusetts Institute of Technology, Cambridge, Massachusetts 02139}
\affiliation{Department of Physics and Astronomy, Shanghai Jiao Tong University, Shanghai, 200240, China}
\author{Karna Morey}
\altaffiliation{Present address: Department of Physics, Stanford University, Stanford, CA 94305}
\affiliation{Department of Physics, Massachusetts Institute of Technology, Cambridge, Massachusetts 02139}
\author{Zongqi Shen}
\affiliation{Department of Physics, Massachusetts Institute of Technology, Cambridge, Massachusetts 02139}
\author{Changmin Lee}
\affiliation{Materials Science Division, Lawrence Berkeley National Laboratory, Berkeley, CA 94720}
\affiliation{Department of Physics, Hanyang University, Seoul 04763, Republic of Korea}
\author{Elizabeth Donoway}
\affiliation{Department of Physics, University of California at Berkeley, Berkeley, CA, USA}
\author{Alex Liebman-Pel\'{a}ez}
\affiliation{Department of Physics, University of California at Berkeley, Berkeley, CA, USA}
\author{Anshul Kogar}
\altaffiliation{Present address: Department of Physics and Astronomy, University of California, Los Angeles, CA, USA}
\affiliation{Department of Physics, Massachusetts Institute of Technology, Cambridge, Massachusetts 02139}
\author{Takashi Kurumaji}
\altaffiliation{Present address: Department of Advanced Materials Science, The University of Tokyo, Kashiwa 277-8561, Japan}
\affiliation{Department of Physics, Massachusetts Institute of Technology, Cambridge, Massachusetts 02139}
\author{Martin Rodriguez-Vega}
\affiliation{Department of Physics, Northeastern University, Boston, MA 02115, USA}
\affiliation{Department of Physics, The University of Texas at Austin, Austin, Texas 78712, USA}
\author{Rodrigo Humberto Aguilera del Toro}
\affiliation{Donostia International Physics Center (DIPC), 20018 Donostia-San Sebasti\'{a}n, Spain}
\affiliation{Centro de F'ısica de Materiales - Materials Physics Center (CFM-MPC), 20018 Donostia-San Sebasti\'{a}n, Spain}
\author{Mikel Arruabarrena}
\affiliation{Centro de F'ısica de Materiales - Materials Physics Center (CFM-MPC), 20018 Donostia-San Sebasti\'{a}n, Spain}
\author{Batyr Ilyas}
\affiliation{Department of Physics, Massachusetts Institute of Technology, Cambridge, Massachusetts 02139}
\author{Tianchuang Luo}
\affiliation{Department of Physics, Massachusetts Institute of Technology, Cambridge, Massachusetts 02139}
\author{Peter M\"{u}ller}
\affiliation{X-Ray Diffraction Facility, Department of Chemistry, Massachusetts Institute of Technology, Cambridge, Massachusetts 02139}
\author{Aritz Leonardo}
\affiliation{Donostia International Physics Center (DIPC), 20018 Donostia-San Sebasti\'{a}n, Spain}
\affiliation{EHU Quantum Center, University of the Basque Country UPV/EHU, 48940 Leioa, Spain}
\author{Andres Ayuela}
\affiliation{Donostia International Physics Center (DIPC), 20018 Donostia-San Sebasti\'{a}n, Spain}
\affiliation{Centro de F'ısica de Materiales - Materials Physics Center (CFM-MPC), 20018 Donostia-San Sebasti\'{a}n, Spain}
\author{Gregory A. Fiete}
\affiliation{Department of Physics, Massachusetts Institute of Technology, Cambridge, Massachusetts 02139}
\affiliation{Department of Physics, Northeastern University, Boston, MA 02115, USA}
\author{Joseph G. Checkelsky}
\affiliation{Department of Physics, Massachusetts Institute of Technology, Cambridge, Massachusetts 02139}
\author{Joseph Orenstein}
\affiliation{Materials Science Division, Lawrence Berkeley National Laboratory, Berkeley, CA 94720}
\affiliation{Department of Physics, University of California at Berkeley, Berkeley, CA, USA}
\author{Nuh Gedik}
\email{gedik@mit.edu}
\affiliation{Department of Physics, Massachusetts Institute of Technology, Cambridge, Massachusetts 02139}

\date{\today}

\maketitle

\subsection{Supporting theoretical discussion}

\subsubsection{Equilibrium strain-spin coupling}\label{sec:equilibriumtheory}

To study the coupling between intrinsic strain and magnetic order in \cmb, theoretical calculations were performed using the \gls{paw} implemented in the \gls{vasp}\cite{kresse_efficient_1996, kresse_from_1999}.
For the exchange and correlation potential, we use the Perdew-Burke-Ernzerhof form of the \gls{gga}, which is further corrected with the Couloumb U parameter following the \gls{gga}+U formulation of Dudarev\cite{dudarev_electron-energy-loss_1998}.
We perform a test calculation of the \gls{pdos} using the HSE06 hybrid functional.
Comparing this with the \gls{gga} and \gls{gga}+U calculations, we determine that for an improved system description, it is necessary to include the terms $U(\ce{Mn})=4$ \si{eV} and $U(\ce{Bi})=3$ \si{eV}.
All calculations were performed with a well-converged plane-wave cutoff energy of $700$ \si{eV}, a gamma-centered $15\times15\times8$ Monkhorst-Pack $k$-point mesh, and a Fermi smearing of $20$ \si{meV}.
Atomic coordinates were relaxed until forces in all directions were smaller than $0.5$ \si{meV/\r{A}}.
An energy convergence criterion of $10^{-7}$ \si{eV} is used.
The atomic valence configuration for Ca, Mn and Bi are 3$s^2$3$p^6$4$s^2$4$p^{0.01}$, 4$s^2$3$d^5$ and 6$s^2$6$p^3$, respectively.
The ground state in the trigonal unit cell shows \gls{afm} order between the Mn atoms with an energy difference of about $200$ \si{meV}.

By including the spin-orbit coupling, we performed additional tests to converge the \gls{mae} with respect to the Brillouin zone sampling.
The spin-orbit coupling energetically favors certain spin orientations in the crystal.
We have determined that the  easy axis is the $x$-axis.
The energy needed to align the spin out-of-plane is $2.4$ \si{meV}, a value that is about a hundred times larger than the \gls{mae} difference with the $y$-axis, which is on the order of $0.02$ \si{meV}.
The \gls{mae} difference with respect  to the $z$-axis remains constant regardless of strain.
The \gls{mae} difference between the $x$ and $y$ directions decreases by applying strain along the $x$-axis, as shown in \cref{fig:equilibriumtheory}.
With a strain of around $0.25 \%$\footnote{Note that, as with any \gls{dft} result, the magnitude of the critical strain from the \gls{gga}+U theory should not be considered quantiatively accurate.}, the easy spin orientation of the crystal changes to lie along the $y$-axis.
 
\subsubsection{Laser-induced dynamics of competing magnetic domains: ``incoherent'' scenario}\label{sec:fluctuations}

In this section, we explore the possibility of interpreting the results of the main text based on the recent quench dynamics theory introduced by Sun and Millis \cite{sun_transient_2020}, and Dolgirev et al. \cite{dolgirev_self-similar_2020}.  

We consider the four observed domains A, B, C, and D as competing AFM orders favored in equilibrium by the local sample anisotropy, regardless of its microscopic origin.
Normally, in the absence of anisotropies, the magnetism in \cmb would be described by model-G \cite{hohenberg1977}.
However, the presence of anisotropies, indicated by the existence of only four domains out of the expected twelve configurations due to symmetry, suggests that total magnetization is not conserved.
Thus, the relaxational behavior of type A studied in detail in \cite{sun_transient_2020,dolgirev_self-similar_2020}, is applicable.
The corresponding relaxational time-dependent-Ginzburg-Landau model is given by $\frac{1}{\gamma_i} \partial_t \psi_i(\mathbf{r}, t)=-\frac{\delta F(t)}{\delta \psi_i(\mathbf{r}, t)}+\eta_i(\mathbf{r}, t),$ where $\psi_i$ corresponds to the staggered magnetization in domain $i=A,B,C,$ and $D.$ $\gamma_i$ and $\eta_i$ are the $i-$th domain relaxation rate and Gaussian white noise source.
The free-energy functional follows the form of Equation (2) in \cite{sun_transient_2020}, with interactions among the domains, $F = \int d^D\mathbf{r} \left(\sum_i f_i+f_c \right)$, $f_i=-\alpha_i \psi_i^2+\left(\xi_{i 0} \nabla \psi_i\right)^2+\psi_i^4$, and $f_c =   \psi^2_C (c_{AC}\psi^2_A+c_{CB}\psi^2_B) + \psi^2_D (c_{AD}\psi^2_A+c_{BD}\psi^2_B)$.
$\xi_{i 0}$ is the bare coherence lengths, and $\alpha_i = \alpha_i(t)$ is a time-dependent dimensionless coefficient influenced by the laser pump excitation's temperature profile.

During the laser pump excitation, characterized by a high temperature and $\alpha_i(t)<0$, and until $\alpha_i(t)=0$, the mean-field order parameters have small values, and the dynamics are governed by order-parameter fluctuations.
Subsequently, the fluctuations grow exponentially, leading to a metastable state determined by the fastest-relaxing domain.
In our experiment, domain A always goes to A, domain B always goes to B, while domain C and D can go to either A or B.
Assuming that the coefficients $\alpha_i$ in equilibrium are equal across the four domains, this observation aligns with the theory prediction if $\gamma_{A/B} > \gamma_{C/D}$ and $\gamma_A \approx \gamma_B$.
 
\subsubsection{Laser-induced dynamics of competing magnetic domains: ``coherent'' scenario}\label{sec:nonequilibriumtheory}

\textit{Group theory aspects of the magnetic order}

\cmb belongs to the trigonal symmorphic space group $P\bar 3m1$ (No. $164$) \cite{gibson_magnetic_2015}.
The Wyckoff positions of the atoms are Bi 2d, Mn 2d, and Ca 1a.
The Ca atoms form a triangular lattice on the top and bottom of the Bi and Mn atoms, while Mn and Bi form  buckled hexagonal lattices.
At the $\Gamma$ point, the point group is $\bar{3}m$, which possess twelve symmetry operations: 
\begin{itemize}
\item[] $\{ 1 | 0 \}$: $(x,y,z) \rightarrow$ $(x,y,z)$
\item[]  $\{ 3^+_{001} | 0 \}$: $(x,y,z) \rightarrow$ $(-y,x-y,z)$
\item[]  $\{  3^-_{001} | 0 \}$: $(x,y,z) \rightarrow$ $(-x+y,-x,z)$
\item[]  $\{ 2_{110} | 0 \}$: $(x,y,z) \rightarrow$ $(y,x,-z)$
\item[]  $\{ 2_{100} | 0 \}$: $(x,y,z) \rightarrow$ $(x-y,-y,-z)$
\item[]  $\{ 2_{010} | 0 \}$: $(x,y,z) \rightarrow$ $(-x,-x+y,-z)$
\end{itemize}
plus their composition with inversion symmetry $\{  -1| 0 \} : (x,y,z) \rightarrow$ $(-x,-y,-z)$.
With this lattice information above the magnetic transition temperature, we proceed with a group theory analysis of the possible magnetic orders with the aid of \small{ISOTROPY} \cite{ISOTROPY}.
In \cmb, the magnetic transition is not accompanied by a lattice deformation, such that the magnetic unit cell coincides with the chemical unit cell.
Therefore, the magnetic order is associated with $\Gamma$ point irreducible representations.
The magnetic subgroups resulting from the onset of all possible $\Gamma$ point irreducible representations are shown in \cref{tab:irreps_mag}.
The $A_{2g}$ and $A_{1u}$ irreducible representations (irreps) correspond to out-of-plane ferromagnetic and antiferromagnetic order, respectively.
The two-dimensional irreps $E_{g}$ and $E_{u}$ induce in-plane magnetic and \gls{afm} order, respectively.
A general direction of the \gls{op} is associated with the magnetic subgroup $P\bar 1'$, which corresponds to the experimental magnetic space group~\cite{gibson_magnetic_2015}.
Notice that special directions of the \gls{op} lead to magnetic structures with higher symmetry and corresponding magnetic space groups $C2/m'$ or $C2'/m$.

Finally, noting that each of the twelve symmetry operations listed above leads to a different \gls{afm} \gls{op} direction (as the real \gls{op} does not lie along any of the special directions), the free energy surface in equilibrium should have exactly twelve minima in the absence of strain, as referenced in the text.

\begin{table}[h]
\begin{tabular}{|l|l|l|l}
\hline
IR                            & Subgroup & Direction  \\ \hline
$\Gamma_1^+$   ($A_{1g}$)   &  $(164.85) P\bar3m1  $       & P1 (a)    \\ \hline
$\Gamma_2^+$   ($A_{2g}$)   &  $ (164.89) P\bar3m'1 $       & P1 (a)   \\ \hline
\multirow{3}{*}{$\Gamma_3^+$ ($E_{g}$)}  &  $ (12.58) C2/m $    & P1 (a,-1.732a) \\ \cline{2-3} 
                              &   $(12.62) C2'/m'$       &   P2 (a,0.577a)        \\ \cline{2-3} 
                              &   $ (2.4) P\bar1$       &     C1  (a,b)     \\ \hline
$\Gamma_1^-$   ($A_{1u}$)   &  $ (164.88) P\bar3'm'1  $       & P1 (a)   \\ \hline
$\Gamma_2^-$   ($A_{2u}$)   &  $(164.87) P\bar3'm1$       & P1 (a)  \\ \hline 
\multirow{3}{*}{$\Gamma_3^-$ ($E_{u}$)}  &  $(12.60) C2'/m $ & P1 (a,-1.732a)   \\ \cline{2-3}    
                              &   $(12.61) C2/m'$       &   P2  (a,0.577a)     \\ \cline{2-3} 
                              &   $ (2.6) P\bar1'$       &     C1  (a,b)     \\ \hline                         
\end{tabular}
\caption{
Subgroups obtained by the onset of magnetic order of a given irreducible representation.
The columns indicate the irreducible representations, the magnetic group, and the \gls{op} direction in the representation space.}
\label{tab:irreps_mag}
\end{table}

\textit{Free energy model}

Now we construct a free energy model considering magnetic order, strain, and phonon modes.
As discussed in the previous section, the in-plane \gls{afm} groundstate corresponds to the magnetic irrep $\Gamma_3^-$, with general in-plane \gls{op} directions $L_1$,  $L_2$.
The parent cell supports  strain mode distortions with irrep $\Gamma_1^+$ and $\Gamma_3^+$.
We consider a two component $\Gamma_3^+$ mode with components $\epsilon_{xx} = - \epsilon_{yy}$ ($\epsilon_1$) and $\epsilon_{xy}$ ($\epsilon_2$).
Finally, the phonon mode ($Q$) is assumed to be totally symmetric.
The free energy, to fourth order, then takes the form $\mathcal F =\mathcal F_M + \mathcal F_S + \mathcal F_C + \mathcal F_{P}$, with
\begin{align}
    \mathcal F_M  & = (T/T_c - 1) L_1^2 + (T/T_c - 1) L_2^2 + (L_1^2 + L_2^2)^2,\\
    \mathcal F_S  & = \frac{1}{2}(e_1^2 + e_2^2)  + e_1^3 -  e_1 e_2^2 + (e_1^2 + e_2^2)^2,\\
    \mathcal F_P  & = \frac{\omega^2}{2}Q^2 +  Q^3 + Q^4 ,\\
    \mathcal F_C  & = \nonumber (L_1^2-L_2^2) e_1 - L_1 L_2 e_2 + (L_1^2+L_2^2)(e^2_1 + e^2_2) \\ \nonumber &+ L_1 L_2 e_1 e_2 + (L_1^2+L_2^2)(Q+Q^2)+(e^2_1 + e^2_2)(Q+Q^2) \nonumber \\ &+(L_1^2 - L_2^2)e_1 Q -L_1 L_2 e_2 Q+e_1^3Q-e_1 e_2^2 Q.
\end{align}
Each of the terms in the model $\mathcal F$ is preceded by a constant whose sign and absolute value cannot be determined within group theory and are sample-dependent.
The magnetic term, $\mathcal F_M$, could include anisotropic terms (due to spin-orbit coupling or single-ion anisotropies, for example), which we omit as the simplest model presented can describe the phenomenology in the experiment. 

We assume that the laser excites the totally-symmetric phonon coherently via a displacive excitation mechanics~\cite{zeiger_theory_1992}.
The laser-phonon coupling term is given by $\mathcal F_l = E_0 \eta(t) Q,$ where $\eta(t) = \kappa \int_0^{\infty} g(t-\tau) e^{-\beta \tau} d \tau,$ and $g(t)$ is the laser pulse shape, and $\beta$ is a parameter associated with rate of electronic relaxation to the groundstate.
The differential equations governing the system can be obtained by taking the variation of the $\mathcal F +\mathcal F_l$ potential with respect to the \gls{afm} \glspl{op}, strain, and phonon.
For the results in the main text, we include phenomenological damping.
The equations of motion are $-\delta (\mathcal F +\mathcal F_l)/\delta \chi_i =  \partial^2_t \chi_i(t)+\gamma_i \partial_t\chi_i(t)$, where  $\chi = \{Q, L_1, L_2\}$ and $\gamma_i$ is the damping coefficient.

\textit{Phonon coupling effect.}

First, we assume that there is no strain in the sample.
The only term linking the \gls{afm} \gls{op} to the time-dependent phonon is $(L_1^2 + L_2^2)(Q+Q^2)$.
We find that this term can lead to flips of $L_1$ and $L_2$ by 180 degrees for long-enough laser pulses. \Cref{fig:nonequilibriumtheory1a} shows the phonon dynamics following photo-excitation with a $1$eV pulse and duration $\tau = 1.1$ \si{ps}.
The phonon frequency was assumed to be $\omega/(2\pi) = 3$ \si{THz}, and the initial temperature is $T=0$.     

\textit{Strain}

The presence of strain mediates coupling terms between the phonon and the \gls{afm} \gls{op} that can lead to orientations absent in equilibrium.
In particular, the coupling term $L_1 L_2 e_2 Q$ induces a solution where the components $L_1, L_2$ oscillate around a position distinct to the initial configuration and to 180-degree flips.
We consider a pulse with duration $\tau = 1.2$ \si{ps}, phonon frequency $\omega/(2\pi) = 3$ \si{THz}, $T=0$, and strains $\epsilon_1 = 0.01$, $\epsilon_1 = 0.05$.
The long-time average of $L_2(t)$ acquires values comparable with $L_1(t)$, as seen in \cref{fig:nonequilibriumtheory2b}. 
 
\subsection{Nonviability of secondary order parameter in describing the \gls{rashg} data}\label{sec:alternativeinterpretation}

Using a combination of \gls{ptmb} and \gls{rashg}, we argue in the main text that, in equilibrium, the magnetism in \cmb that we observe is described by a spatial distribution of four domains $A$, $B$, $C$, and $D$, for which the order parameters satisfy the relations $l_A = -l_B$, $l_C = -l_D$, and $l_C = R(l_A)$, where $R$ is an element of the high-temperature point group \htpg.
In this scenario, the contrast in SHG between e.g. $l_A$ and $l_B$ comes from the fact that there can be two contributions to the \gls{rashg} signal which interfere with each other and transform differently as $l_A \rightarrow l_B$.
That is, we can write
\begin{equation}
I_\mathrm{SHG} \propto |e^\mathrm{out}_i \chi_{ijk} e^\mathrm{in}_j e^\mathrm{in}_k|^2,
\end{equation}
where
\begin{equation}
\begin{aligned}
\chi_{ijk} &\equiv &&\chione_{ijk} + \chitwo_{ijk}, \\
\end{aligned}
\end{equation}
\begin{equation}
\label{eq:trrelations}
\begin{aligned}
T (\chione_{ijk}) &= &-&\chione_{ijk}, \\
T (\chitwo_{ijk}) &= &&\chitwo_{ijk},
\end{aligned}
\end{equation}
$T$ is the time-reversal operator which takes $l_A \rightarrow -l_A = l_B$, and $e^\mathrm{in}$ and $e^\mathrm{out}$ are unit vectors in the direction of the polarization of the incoming and outgoing electric fields, respectively.

An important question thus arises: what is the origin of the two sources $\chione$ and $\chitwo$?
The simplest explanation is that both contributions are due to the \gls{afm} \gls{op}; the relations in \cref{eq:trrelations} are then permitted below $T_c$, where time-reversal symmetry is broken and so $\chi_{ijk}$ can, in general, have parts which transform as both even and odd under time-reversal.
However, an alternative interpretation is that there are in fact two \glspl{op} $l$ and $g$, with $g$ being e.g. some structural \gls{op}, so that, e.g.
\begin{equation}
\label{eq:proptolg}
\begin{aligned}
\chi^C &\propto l + g, \text{and} \\
\chi^D &\propto l + Rg,
\end{aligned}
\end{equation}
for some $R \in \htpgmathmode$.

In this interpretation, alternative descriptions of the time-resolved results, involving e.g. a melting of $g$ while the $l$ stays fixed, may explain \cref{fig2}, but not imply a reorientation of $l$ as we claim in the main text.
However, we argue that such an interpretation is not a viable description of our results, both in and out of equilibrium.
This can be understood by considering the fact that the \gls{rashg} pattern in \cref{fig1:f} has nodes at $\phi=0$ and $\phi=\pi$ for all temperatures below $T_c$ (see \supsupcref{ctempdep}).
In the case $Rg = -g$, this suggests that, if there are indeed two independent order parameters contributing to the \gls{rashg} pattern, their amplitudes must exactly cancel for all $T < T_c$.
This is unlikely as the two independent order parameters should in general depend differently on temperature.
Moreover, the case $Rg \neq -g$ is difficult to reconcile in light of the result that the \gls{ptmb} signal is invariant under thermal cycling.
In nonequilibrium, any scenario involving the melting of one \gls{op} while the other stays fixed must also be reconciled with the observation that the nodes in \cref{fig2:c} in the initial state (before $t=0$) become peaks in the final state.

Furthermore, the above interpretation must also explain how the final state of the transition seems to differ from sample to sample (as explained in the main text; see \supsupcref{CtoBshort,CtoBlong}), as well as how the \gls{rashg} is poled in equilibrium to only two of many different degenerate states.
Both of these facts fall out naturally from the assignment presented in the main text.

Together with the lack of a change in the \gls{xrd} across $T_c$ (see \supsupcref{xrd}), the above considerations suggest that the scenario presented in the main text, which neatly describes the data both in and out of equilibrium, is the most likely.
Future studies will be required to completely truly rule out all alternative scenarios.
 
\subsection{Fits to susceptibility tensors}\label{sec:fitting}

In the text, it was remarked that the \gls{op} assignment depicted in \crefrange{fig1}{fig2} was consistent with the \gls{shg} susceptibility tensors extracted by fitting the \gls{rashg} patterns.
In this section, we state how this fitting was performed.

Before that discussion, however, we must mention one important point about the predictive power of these fits.
Since the low-temperature point group of \cmb is $\bar{1}'$, the relevant susceptibility tensors are entirely unconstrained and the cost function to be minimized may involve upward of 36 free parameters.\footnote{\label{foot:36parameters}
Here, the number 36 comes from observing that $\chi^e_{ijk}$, the second-order susceptibility tensor for electric-dipole \gls{shg} (which is the simplest \gls{shg} source allowed in the magnetic point group of \cmb), nominally has 27 elements, each of which may have a real or imaginary part, resulting in 54 free parameters.
However, due to the intrinsic $j\leftrightarrow k$ symmetry in the response function
\[
P_i(2\omega) = \chi_{ijk} E_j(\omega)E_k(\omega),
\]
many of these elements are constrained to be equivalent, resulting ultimately in 36 free parameters.
}
Thus, in principle, many combinations of parameters may fit the data appropriately, thereby limiting the extent to which the \gls{shg} fits can be used to draw conclusions about the underlying state.
One must look to other arguments, e.g. involving \gls{ptmb}, to draw useful conclusions about the data, as was done in the text.
However, once a conclusion has been made, it can at least be checked that a set of fitting parameters \textit{does} exist that is consistent with the conclusion and with the data.
This is the goal of this section.

The \gls{op} assignment discussed in the text involves two parts.
In the first part, the equilibrium SHG in the four domains of \cref{fig1} is assigned to four different \gls{op} directions, in which the \gls{shg} on consecutive cooldowns on a single spot is due to $180^\circ$ \gls{afm} configurations, and adjacent spots have a relative angle between them.
As argued in the text, since the $180^\circ$ domains have different \gls{rashg} patterns, there are most likely two contributions to the \gls{shg} adding coherently to give an interference term, as observed in other magnetic systems\cite{fiebig_second_1994, fiebig_second_2001, fiebig_second-harmonic_2005}.
Designating the four configurations as $A$, $B$, $C$, and $D$ (for \cref{fig1:d}, \cref{fig1:e}, \cref{fig1:f}, and \cref{fig1:g}, respectively), we thus have
\begin{equation}
\begin{aligned}
\label{eq:equilibrium}
\chi^{A}_{ijk} &= &&&&\chione_{ijk}+\chitwo_{ijk}\\
\chi^{B}_{ijk} &= &&&-&\chione_{ijk}+\chitwo_{ijk}\\
\chi^{C}_{ijk} &= &&R(&&\chione_{ijk}+\chitwo_{ijk})\\
\chi^{D}_{ijk} &= &&R(&-&\chione_{ijk}+\chitwo_{ijk}),
\end{aligned}
\end{equation}
where $\chione_{ijk}$ and $\chitwo_{ijk}$ are the susceptibility tensors corresponding to the two components discussed above, and $R(\cdot)$ refers to an operation by some element $R$ of the high-temperature point group ($\bar{3}m$), i.e.
\begin{equation}
R(\chi_{ijk}) \equiv R_{il}R_{jm}R_{kn}\chi_{lmn}.
\end{equation}

\begin{table*}
\renewcommand{\arraystretch}{1.1}
\caption[Vector definition of polarization channels.]{\label{table:efielddefinitions}Vector definition of polarization channels. $\theta=10^\circ$ is the angle of incidence and $\phi$ is the angle of the plane of incidence with respect to the $\hat{x}$ axis (see \cref{fig0:c}).}
\begin{tabular}{|c c|c c c|c c c|c c c|}
\hline
\multirow{2}{*}{Input} & \multirow{2}{*}{Output} & \multicolumn{3}{|c|}{$\hat{e}^\mathrm{in}(\phi)$} & \multicolumn{3}{|c|}{$\hat{e}^\mathrm{out}(\phi)$} & \multicolumn{3}{|c|}{$\hat{k}^(\phi)$}\\
& & $\hat{e}^\mathrm{in}_x(\phi)$ & $\hat{e}^\mathrm{in}_y(\phi)$ & $\hat{e}^\mathrm{in}_z(\phi)$ & $\hat{e}^\mathrm{out}_x(\phi)$ & $\hat{e}^\mathrm{out}_y(\phi)$ & $\hat{e}^\mathrm{out}_z(\phi)$& $\hat{k}_x(\phi)$ & $\hat{k}_y(\phi)$ & $\hat{k}_z(\phi)$\\
\hline
P & P & $\cos\phi\cos\theta$ & $\sin\phi\cos\theta$ & $-\sin\theta$ & $-\cos\phi\cos\theta$ & $-\sin\phi\cos\theta$ & $-\sin\theta$ & \multirow{4}{*}{$-\cos\phi\sin\theta$} & \multirow{4}{*}{$-\sin\phi\sin\theta$} & \multirow{4}{*}{$-\cos\theta$} \\
P & S & $\cos\phi\cos\theta$ & $\sin\phi\cos\theta$ & $-\sin\theta$ & $-\sin\phi$ & $\cos\phi$ & $0$ & & &\\
S & P & $-\sin\phi$ & $\cos\phi$ & $0$ & $-\cos\phi\cos\theta$ & $-\sin\phi\cos\theta$ & $-\sin\theta$& & &\\
S & S & $-\sin\phi$ & $\cos\phi$ & $0$ & $-\sin\phi$ & $\cos\phi$ & $0$& & &\\
\hline
\end{tabular}
\end{table*}

In the time-resolved case, \cref{eq:equilibrium} should hold before zero delay, and a similar set of equations should hold at long times in the final state:
\begin{equation}
\begin{aligned}
\label{eq:nonequilibrium}
\chi^{A'}_{ijk} &= &&\alpha\chione_{ijk}+\chitwo_{ijk}\\
\chi^{B'}_{ijk} &= &-&\alpha\chione_{ijk}+\chitwo_{ijk}\\
\chi^{C'}_{ijk} &= &&\alpha\chione_{ijk}+\chitwo_{ijk}\\
\chi^{D'}_{ijk} &= &&\alpha\chione_{ijk}+\chitwo_{ijk},
\end{aligned}
\end{equation}
i.e., $A\rightarrow A'$, $B\rightarrow B'$, $C\rightarrow A'$, and $D\rightarrow A'$, as in \cref{fig2}.
Here, $\alpha$ is an overall factor to account for the fact that the magnitude of the \gls{op} may change due to heating following the transition.
We also allow a relative scale $\beta$ (close to unity) between $A$ / $B$ / $A'$ / $B'$ and $C$ / $D$ / $C'$ / $D'$ to account for the fact that these are taken on different spots on the sample.
For $\chione_{ijk}$, we use the totally asymmetric complex electric dipole \gls{shg} tensor in the point group $\bar{1}'$, which has $36$ independent elements (see \cref{foot:36parameters}).
In the language of \onlinecref{birss}, this would be referred to as a $c$-type tensor and should have purely imaginary components; however, in the presence of dispersion or dissipation, complex components are also allowed.
For $\chitwo_{ijk}$, many options are possible, including an $i$-type electric dipole component, electric quadrupole component, or magnetic dipole component, each of which may couple with the \gls{afm} \gls{op}.
Note that $\chitwo_{ijk}$ is independent of $\phi$ for electric dipole \gls{shg}, but for magnetic dipole or electric quadrupole \gls{shg}, it acquires a dependence on $\phi$ due to a factor of the wavevector $\vec{k}$ of the incident light (defined in \cref{table:efielddefinitions}).
From \onlinecref{kumar_magnetic_2017}, for electric dipole \gls{shg} we have
\begin{equation}
\chitwo_{ijk}(\phi) = \chi^e_{ijk},
\end{equation}
whereas for electric quadrupole \gls{shg} we have
\begin{equation}
\chitwo_{ijk}(\phi) = i\chi^q_{lijk}k_l(\phi),
\end{equation}
and for magnetic dipole \gls{shg} we have
\begin{equation}
\chitwo_{ijk}(\phi) = i\epsilon_{ilm}\chi^m_{mjk}k_l(\phi),
\end{equation}
where $\epsilon$ is the Levi-Civita tensor, and $\chi^e, \chi^q$, and $\chi^m$ are susceptibility tensors (in the low-temperature magnetic point group $\bar{1}'$) defined by the response equations
\begin{equation}
P_i(2\omega) = \chi^e_{ijk}E_j(\omega)E_k(\omega),
\end{equation}
\begin{equation}
Q_{ij}(2\omega) = \chi^q_{ijkl}E_k(\omega)E_l(\omega),
\end{equation}
and
\begin{equation}
M_i(2\omega) = \chi^m_{ijk}E_j(\omega)E_k(\omega).
\end{equation}

To begin, let us rotate the system so that the optical axis (the $z$ axis in \cref{fig0:c}) is normal to the $(hkl)=(011)$ plane.
This amounts to taking
\begin{equation}
\label{eq:orient_eee}
\chi^e_{ijk} \rightarrow \chi^{e'}_{ijk}=O_{il}O_{jm}O_{kn}\chi^e_{lmn}
\end{equation}
\begin{equation}
\label{eq:orient_mee}
\chi^m_{ijk} \rightarrow \chi^{m'}_{ijk}=O_{il}O_{jm}O_{kn}\chi^m_{lmn}
\end{equation}
\begin{equation}
\label{eq:orient_qee}
\chi^q_{lijk} \rightarrow \chi^{q'}_{lijk}=O_{lm}O_{in}O_{jp}O_{kq}\chi^q_{mnpq},
\end{equation}
where\footnote{
The orientation matrix $O$ was calculated using VESTA\cite{vesta}.
}
\begin{equation}
O \equiv \left(\begin{matrix}
&1 &\phantom{-}0 &\phantom{-}0\\
&0 &\phantom{-}0.460230  &-0.887799\\
&0 &\phantom{-}0.887799  &\phantom{-}0.460230
\end{matrix}\right)
\end{equation}
in \cmb.
Then, for each choice of $R$ and of the source (elecric dipole, electric quadrupole, or magnetic dipole) we choose for $\chitwo$, we simultaneously fit the four \gls{rashg} polarization combinations $p \in \{\shortPP, \shortPS, \shortSP, \shortSS\}$ at $\Delta t=-2$ \si{ps} ($d\in \{A, B, C, D\}$) and at $\Delta t\approx 40$ \si{ps} ($d \in \{A', B', C', D'\}$) to the following model:
\begin{equation}
\label{eq:mastermodel}
\begin{aligned}
I_\mathrm{SHG}^{p,d}&(\phi, \psi, \{\chione\}, \{\chitwo\}, \alpha, \beta)\\
 = &|\chi'^{d}_{ijk}(\phi-\psi, \{\chione\}, \{\chitwo\}, \alpha, \beta)\\
\cdot & \hat{e}_i^{p,\mathrm{out}}(\phi-\psi)\hat{e}^{p, \mathrm{in}}_j(\phi-\psi)\hat{e}_k^{p,\mathrm{in}}(\phi-\psi)|^2,
\end{aligned}
\end{equation}
where $\{\chione\}$ and $\{\chitwo\}$ are the free parameters of $\chione_{ijk}$ and $\chitwo_{ijk}$, and $\hat{e}^{p,\mathrm{out}}$ and $\hat{e}^{p,\mathrm{in}}$ are defined in \cref{table:efielddefinitions}.
For each source (electric dipole, electric quadrupole, and magnetic dipole) of $\chitwo$, is is found that there exists $R\in D_{3d}$ such that \cref{eq:mastermodel} produces an appropriate fit to the data (see \cref{fitting}). 
 
\subsection{Error bar estimation for \cref{fig3:a,fluencedep}\label{sec:errorbars}}

The errorbars in \cref{fig3:a,fluencedep} are ultimately due to uncertainty in the value of the \gls{shg} intensity at each angle $\phi$.
When these angles are integrated over, the associated uncertainties in the \gls{shg} intensity are summed in quadrature.
To extract the uncertainties in the \gls{shg} intensity as a function of angle is a somewhat difficult process, as these uncertainties are largely systematic rather than statistical.
Indeed, the \gls{rashg} pattern should in general be a smooth function of $\phi$; however, it is seen (e.g. \cref{fig2:d}) that the \gls{rashg} patterns are not necessarily smooth.
We therefore estimate the size of the errorbars using more heuristic methods, which, while imprecise, should at least give a rough approximation to the true uncertainty.

These heuristic errorbars in the \gls{shg} intensity are found by estimating the size of the variations in the \gls{rashg} pattern compared to a smoothed approximation (approximated with an $N=6$ FFT filter).
The errorbars corresponding to the data in \supcref{fluencedep} are shown in \cref{errorbars}, and the errorbars in \cref{fig3:a} are computed with an identical procedure on that dataset.
 
\subsection{Orientation of the \cmb crystals by fitting the high-temperature \glsfmtshort{rashg} data}\label{sec:orientation}

While the surface normal direction of the \cmb crystal used in \cref{fig0,fig1,fig2,quenchtimes,fig3} is known to be (011) (the natural cleavage plane of the material), the azimuthal angles (in the lab frame) corresponding to the projections of the $a$, $b$, and $c$ axes in this plane are not known \textit{a priori}.
To determine these angles, we fit the \gls{rashg} data in the four polarization channels at high temperature (\supcref{fullequilibrium:a}) to both electric quadrupole and magnetic dipole models (which are the only two \gls{shg} sources allowed at high temperature), whose susceptibility tensors are defined in the high temperature point group $D_{3d}$.
The model function for the electric quadrupole fit was
\begin{equation}
\begin{aligned}
I_\mathrm{SHG}^p(\phi, \psi, \{\chithree\}) = |&\chithreeprime_{ijk}(\phi-\psi, \{\chithree\})\\
&\cdot \hat{e}_i^{p,\mathrm{out}}(\phi-\psi)\hat{e}_j^{p,\mathrm{in}}(\phi-\psi)\hat{e}_k^{p,\mathrm{in}}(\phi-\psi)|^2,
\end{aligned}
\end{equation}
and the model function for the magnetic dipole fit was
\begin{equation}
\begin{aligned}
I_\mathrm{SHG}^p(\phi, \psi, \{\chifour\}) = |&\chifourprime_{ijk}(\phi-\psi, \{\chifour\})\\
&\cdot \hat{e}_i^{p,\mathrm{out}}(\phi-\psi)\hat{e}_j^{p,\mathrm{in}}(\phi-\psi)\hat{e}_k^{p,\mathrm{in}}(\phi-\psi)|^2,
\end{aligned}
\end{equation}
where $\hat{e}^{p,\mathrm{in}}$ and $\hat{e}^{p,\mathrm{out}}$ are defined in \cref{table:efielddefinitions}, $\chithreeprime_{ijk}$ is given by
\begin{equation}
\chithreeprime_{ijk}(\phi) = iO_{lm}O_{in}O_{jp}O_{kq}\chithree_{mnpq}k_l(\phi)
\end{equation}
with $\vec{k}$ defined in \cref{table:efielddefinitions}, $\chifourprime_{ijk}$ is given by
\begin{equation}
\chifourprime_{ijk}(\phi) = i\epsilon_{ilm} O_{mn}O_{jq}O_{kr}\chifour_{nqr}k_l(\phi),
\end{equation}
and $\chithree_{mnpq}$ and $\chifour_{nqr}$ are the susceptibility tensors for electric quadrupole and magnetic dipole \gls{shg}, respectively, constrained by the high temperature point group $D_{3d}$ (and with free parameters $\{\chithree\}$ and $\{\chifour\}$).
Here, $O$ (defined in \cref{sec:fitting}) is the orientation matrix which projects the crystal along the vector normal to (011), and rotates it so that the projection of the $a$ axis on this plane is purely along $x$.
$\phi'\equiv\phi-\psi$ is thus defined so that $\phi'=\ang{0}$ corresponds to the angle at which the plane of incidence is also parallel to that axis.

A good fit to the data (see \supcref{fig:orientation}) is found for $\psi=\psi_m\equiv\ang{73.7+-3.1}$ (magnetic dipole fit) or $\psi=\psi_q\equiv\ang{78.3+-2.1}$ (electric quadrupole fit).
Since $\psi_m$ and $\psi_q$ differ by only a few degrees, for illustrative purposes it suffices to use the approximate angle $\psi_0\equiv(\psi_m+\psi_q)/2=\ang{76.0}$; all of the raw data collected from the experimental apparatus was therefore shifted in post-processing by $\psi_0$ to account for this angle.
 
\subsection{Evidence for the interference of two \glsfmtshort{shg} contributions in the \glsfmtshort{rashg} data\label{sec:interference}}

In the main text, we argued that the \ang{180}-opposite \gls{afm} domains in \cmb are observable in \gls{rashg} because of an interference between two sources of \gls{shg} which transform differently under inversion.
A necessary condition for this is of course that there are two SHG sources which interfere coherently below the magnetic ordering temperature.
Here, we describe evidence from the temperature dependence of the \gls{rashg} patterns that confirms the existence of multiple interfering sources of \gls{shg} below $T_c$ in \cmb.

In \cref{interference}, we plot the temperature dependence of the \gls{rashg} pattern in the domain corresponding to \cref{fig1:f}.
It is observed that, at some angles, the \gls{shg} intensity decreases rather than increasing when the magnetic order sets in at $T_c$.
Assuming that the amplitude of the \gls{shg} source $\vec{P_M}$ corresponding to the magnetic order is a monotonic function of the \gls{op}, this is direct evidence of the existence of two \glsfmtshort{shg} sources, $P_M$ and $P_0$, which are interfering below $T_c$.
To see this, we write the total \gls{shg} intensity as the squared absolute value of the sum of $P_M$ and $P_0$\citep{fiebig_second-harmonic_2005}:
\begin{equation}
\begin{aligned}
I_\mathrm{SHG} &\propto |P_M + P_0|^2
& \propto |P_M|^2 + |P_0|^2 + 2|P_M||P_0|\Delta,
\end{aligned}
\end{equation}
where $\Delta$ characterizes the extent to which $P_M$ and $P_0$ interfere coherently.
Constructive or destructive interference occurs when $\Delta \neq 0$.

Assuming $|P_M|$ and $|P_0|$ are monotonic functions of the \gls{op} (which should be true at least in the vicinity of the phase transition) and that the \gls{op} itself is monotonic as a function of temperature (which is known via neutron diffraction\citep{gibson_magnetic_2015}), a decreasing value of $I_\mathrm{SHG}$ is only possible when $\Delta < 0$.
\Cref{interference} shows a plot of this occuring in the temperature dependence of the \gls{shg} intensity corresponding to \cref{fig1:f}.
Thus, the \gls{rashg} patterns presented in this work necessarily involve two simultaneous contributions to the total \gls{shg} intensity which interfere coherently.
 
\subsection{Origin of the \glsfmtshort{ptmb} signal above $T_c$\label{sec:hightempptmb}}

The non-zero birefringence above the transition temperature is due to the crystallographic point group of the crystal ($D_{3d}$) coupled with the orientation of the crystal in the experiment (the upward-facing crystal direction is $(011)$).
To see this, note that the birefringence signal comes from a difference between the $xx$ and $yy$ components of the dielectric permittivity tensor $\epsilon_{ij}$.
In the $D_{3d}$ point group, $\epsilon_{ij}$ takes the form
\begin{equation}
\epsilon_{ij} = \left(\begin{matrix}
a & 0 & 0 \\
0 & a & 0 \\
0 & 0 & c
\end{matrix}\right)_{ij}
\end{equation}
in the reference frame where the $(001)$ axis is along $z$.
In the lab frame, it is the $(011)$ axis which is directed along $z$, so that the same tensor in the lab frame takes the form
\begin{equation}
\begin{aligned}
\epsilon'_{ij} &= O_{ik}O_{jl}\epsilon_{kl} \\
&= \left(\begin{matrix}
a & 0 & 0 \\
0 & 0.211812a+0.788188c & 0.408592a-0.408592c \\
0 & 0.408592a-0.408592c & 0.788188a+0.211812c
\end{matrix}\right)_{ij}
\end{aligned}
\end{equation}
where
\begin{equation}
O_{ij} = \left(\begin{matrix}
1 & 0 & 0 \\
0 & 0.460230 & -0.887799 \\
0 & 0.887799 & 0.460230
\end{matrix}\right)_{ij}
\end{equation}
(see \cref{sec:fitting}).
Thus, in the lab frame, $\epsilon'_{xx}$ and $\epsilon'_{yy}$ are different, which results in a nonzero birefringence signal.
Put differently, it is the fact that the crystal is oriented along $(011)$ in our experiment which breaks the nominal rotational symmetry afforded by the $D_{3d}$ point group above the transition temperature.
Owing to the exceptional signal-to-noise ratio enabled by the combination of balanced photodetection and lock-in detection methods, the PTMB measurements are sensitive to extremely small changes ($\sim$\qty{1e-6}-\qty{1e-7}) in this signal, enabling observation of minuscule changes in the lattice constant (due to, e.g., thermal contraction) that are not detectable through XRD.

\begin{widetext}
\clearpage
\begin{figure}
\centering
\includegraphics[width=180mm]{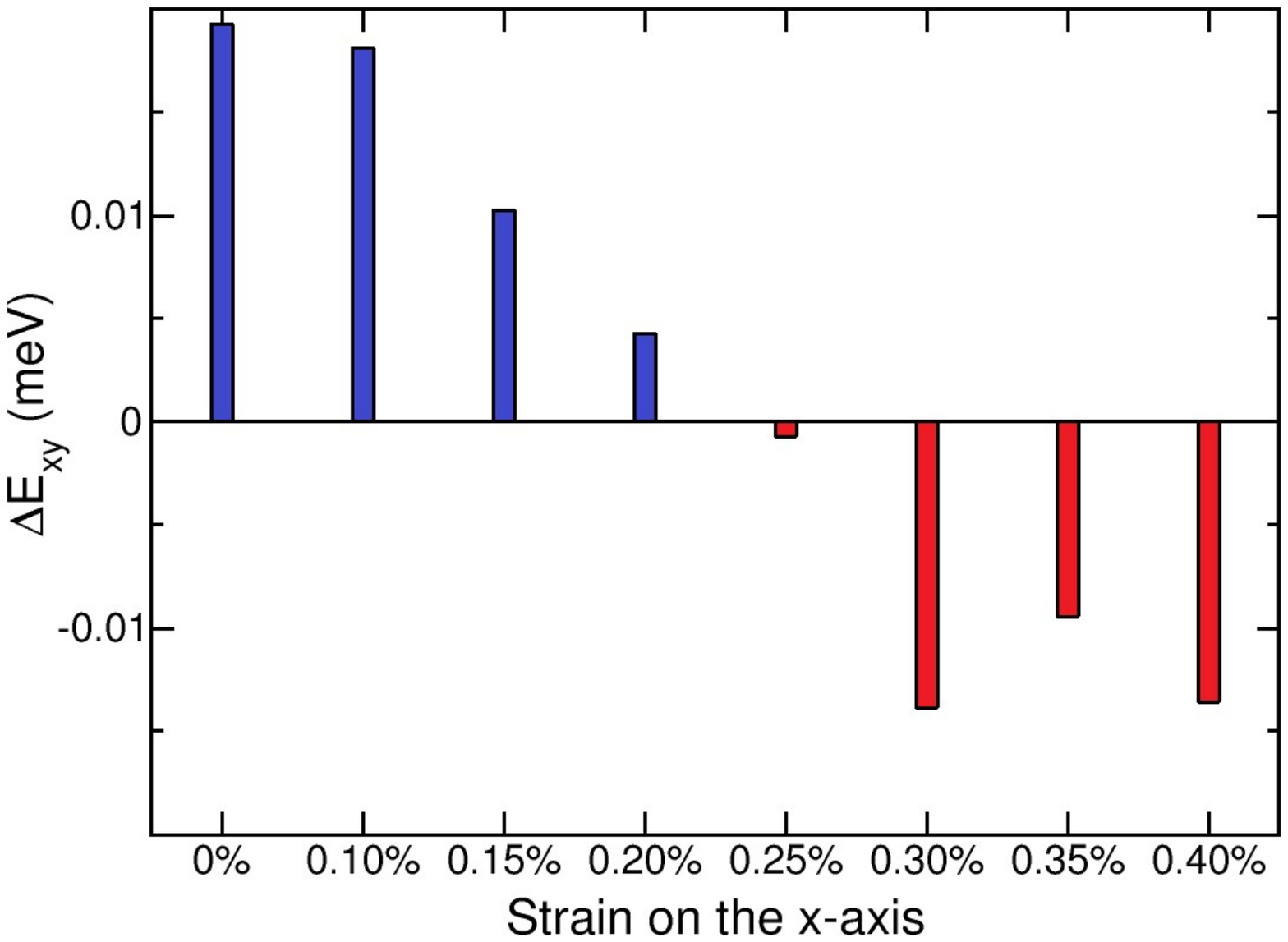}
\caption{
\label{fig:equilibriumtheory}
Energy difference between $x$ and $y$-directions with respect to the easy axis.
Blue bars indicate the $x$-direction for the \glsfmtshort{afm} \glsfmtshort{op} is preferred, and red bars indicate the $y$-direction for the \glsfmtshort{afm} \glsfmtshort{op} is preferred.
Note that for strains in experiments, the difference value goes to zero, indicating a change in the in-plane magnetic moment direction.
}
\end{figure}
 
\clearpage
\begin{figure}
\phantomsubfloat{\label{fig:nonequilibriumtheory1a}}
\phantomsubfloat{\label{fig:nonequilibriumtheory1b}}
\centering
\includegraphics[width=180mm]{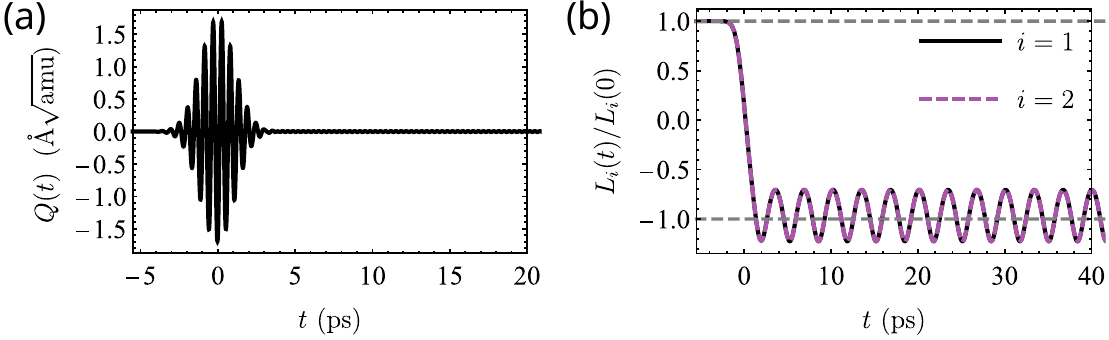}
\captionsetup{singlelinecheck=off}
\caption[]{
\label{fig:nonequilibriumtheory1}
\begin{enumerate*}[label=\caplabel, ref=\capref]
\item Phonon mode and \item \glsfmtshort{afm} \glsfmtshort{op} following photo-excitation with coupling $(L_1^2 + L_2^2)(Q+Q^2)$.
\end{enumerate*}
}
\end{figure}
 
\clearpage
\begin{figure}
\centering{
\phantomsubfloat{\label{fig:nonequilibriumtheory2a}}
\phantomsubfloat{\label{fig:nonequilibriumtheory2b}}
}
\includegraphics[width=180mm]{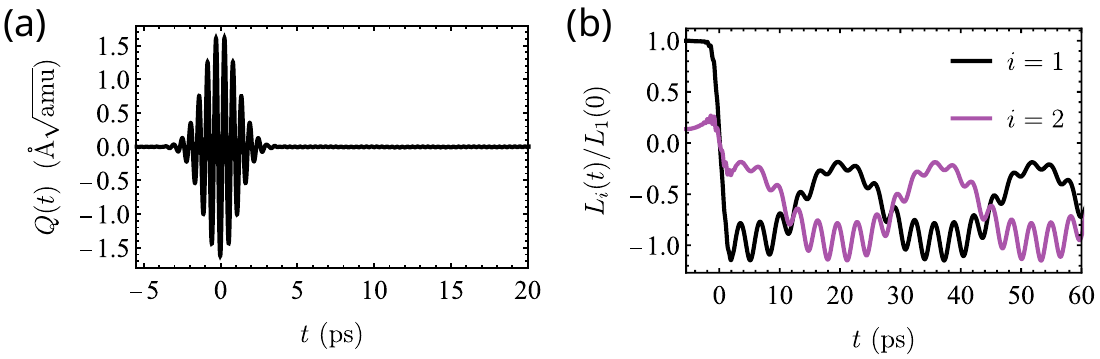}
\captionsetup{singlelinecheck=off}
\caption[]{
\label{fig:nonequilibriumtheory2}
\begin{enumerate*}[label=\caplabel, ref=\capref]
\item Phonon mode and \item \glsfmtshort{afm} \glsfmtshort{op} following photo-excitation with coupling $L_1 L_2 e_2 Q$.
\end{enumerate*}
}
\end{figure}
 
\clearpage
\begin{figure}
\centering
\includegraphics{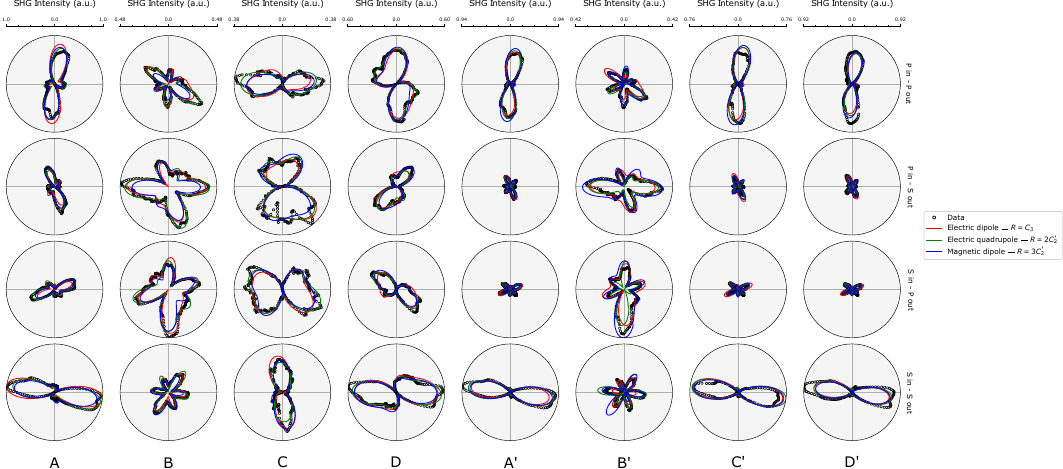}
\caption{
\label{fitting}
Fits to the \glsfmtshort{rashg} pattern in the four domains of \cref{fig1} $2.0$ \si{ps} before ($A$, $B$, $C$, $D$) and $\apx 40$ \si{ps} after ($A'$, $B'$, $C'$, $D'$) zero delay.
Data is represented by black circles; solid lines are fits to \seqs{1}{2} with $\chitwo$ denoted in the legend.
}
\end{figure}
 
\clearpage
\begin{figure}
\centering{
\phantomsubfloat{\label{fluencedep:a}}
\phantomsubfloat{\label{fluencedep:b}}
}
\includegraphics{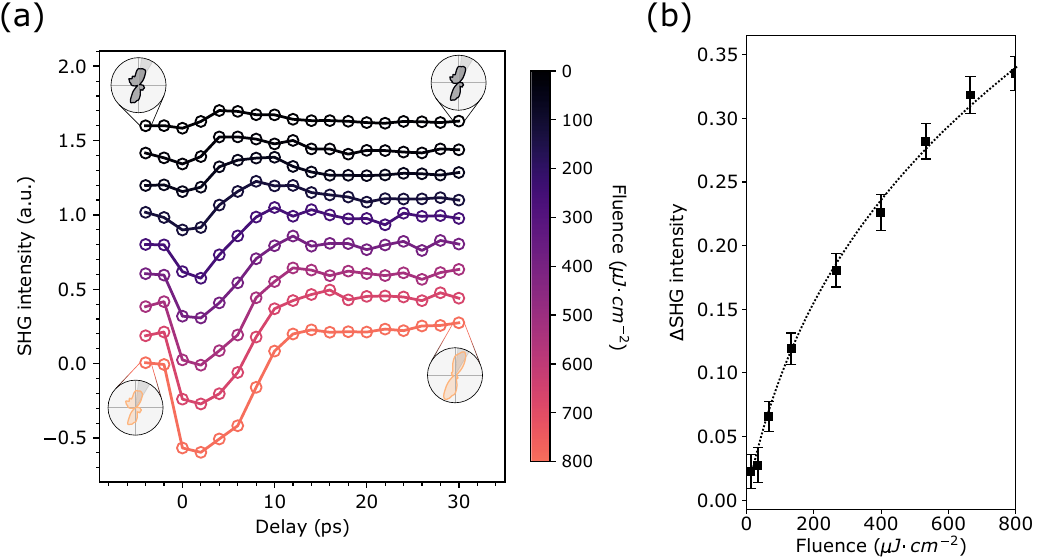}
\captionsetup{singlelinecheck=off}
\caption[]{
\label{fluencedep}
\begin{enumerate*}[label=\caplabel, ref=\capref]
\item Integrated SHG intensity for a representative \glsfmtshort{afm} domain as a function of delay and pump fluence.
Insets show the initial (left) and final (right) states at high (bottom) and low (top) pump fluences, as well as the integration region (indicated by the shaded area).
\item Difference between the intensity of the integration region specified in \ref{fluencedep:a} at long times (averaged from $\Delta t=40$ to $\Delta t=50$ \si{ps}) and the intensity before zero delay (averaged from $\Delta t=-4$ to $\Delta t=-2$ \si{ps}).
Dashed line is a guide to the eye.
\end{enumerate*}
}
\end{figure}
 
\clearpage
\begin{figure}
\centering
\includegraphics{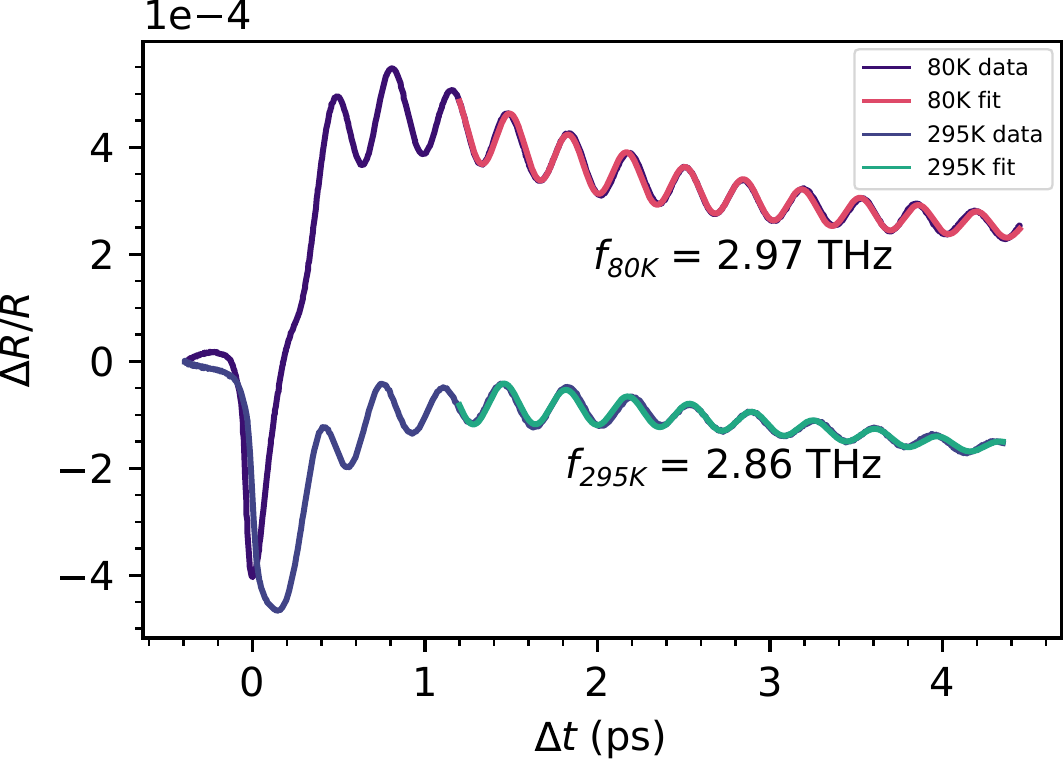}
\captionsetup{singlelinecheck=off}
\caption[]{
\label{pp}
Time-resolved reflectivity obtained from \cmb at $80$ \si{K} and $295$ \si{K}.
The pump and probe wavelengths were both $800$ \si{nm}, the pulse width was $28$ \si{fs}, and the pump and probe fluences were $10$ and $1$ \si{\mu J \cdot cm^{-2}}, respectively.
The data is windowed after the initial transients and then fit to a damped harmonic oscillator plus a polynomial background.
}
\end{figure}
 
\clearpage
\begin{figure}
\centering
\includegraphics{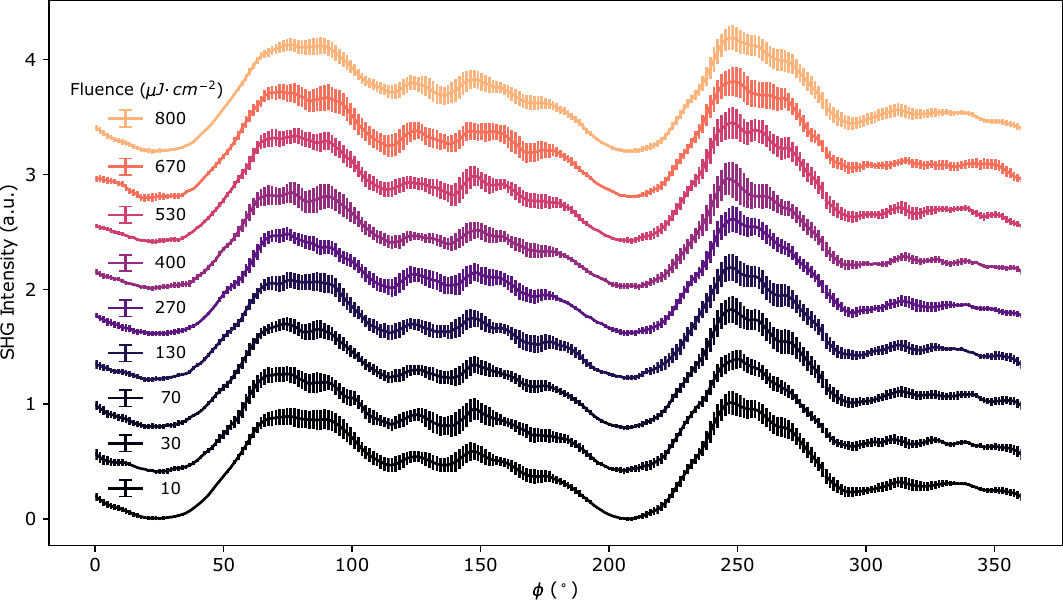}
\caption{\label{errorbars}\glsfmtshort{rashg} results of \supcref{fluencedep} depicted with estimated errorbars (see \supcref{sec:errorbars}).}
\end{figure}
 
\clearpage
\begin{figure}
\centering
\includegraphics{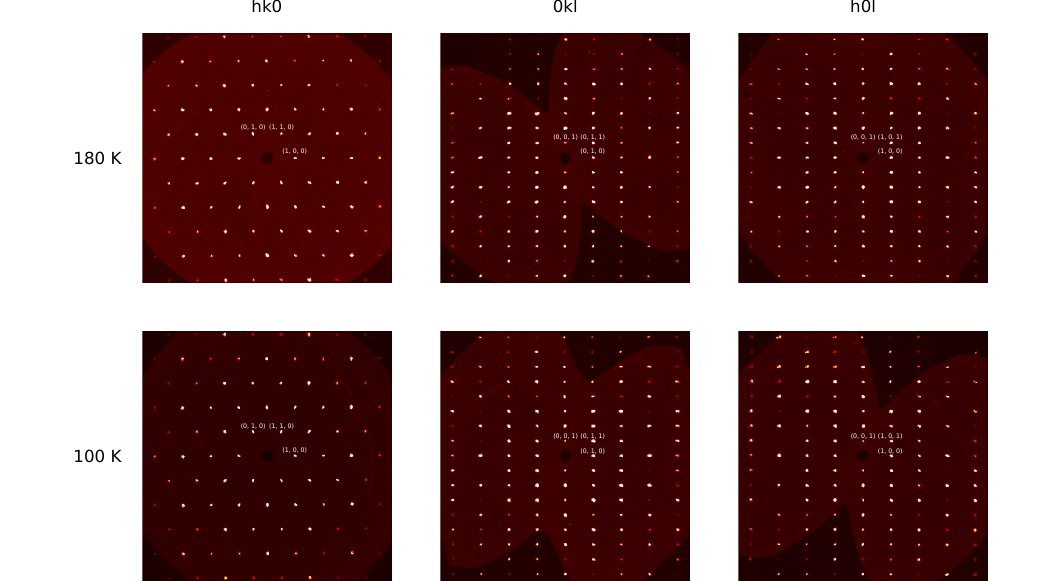}
\captionsetup{singlelinecheck=off}
\caption[]{
\label{xrd}
Single-crystal \glsfmtshort{xrd} precession images obtained from \cmb.
The high temperature refinement is in agreement with previous reports\cite{gibson_magnetic_2015}.
No change in the crystal structure is observed across $T_c=150$ \si{K}.
}
\end{figure}
 
\clearpage
\begin{figure}
\centering
\includegraphics{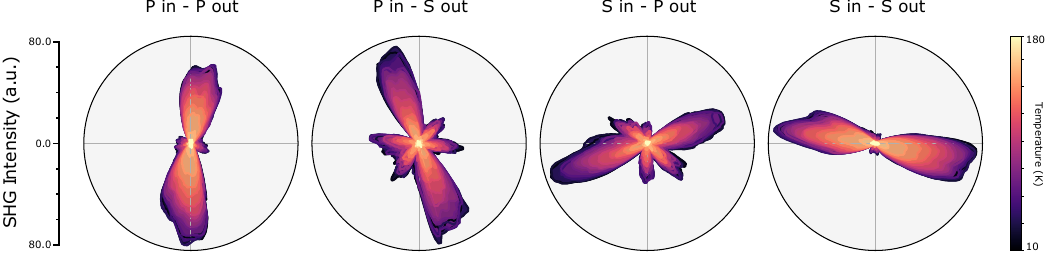}
\caption{\label{fulltempdep}\glsfmtshort{rashg} results of \cref{fig0} depicted in all four polarization channels (\PP, \PS, \SP, and \SS).}
\end{figure}
 
\clearpage
\begin{figure}
\centering{
\phantomsubfloat{\label{fullequilibrium:a}}
\phantomsubfloat{\label{fullequilibrium:b}}
\phantomsubfloat{\label{fullequilibrium:c}}
\phantomsubfloat{\label{fullequilibrium:d}}
\phantomsubfloat{\label{fullequilibrium:e}}
}
\includegraphics{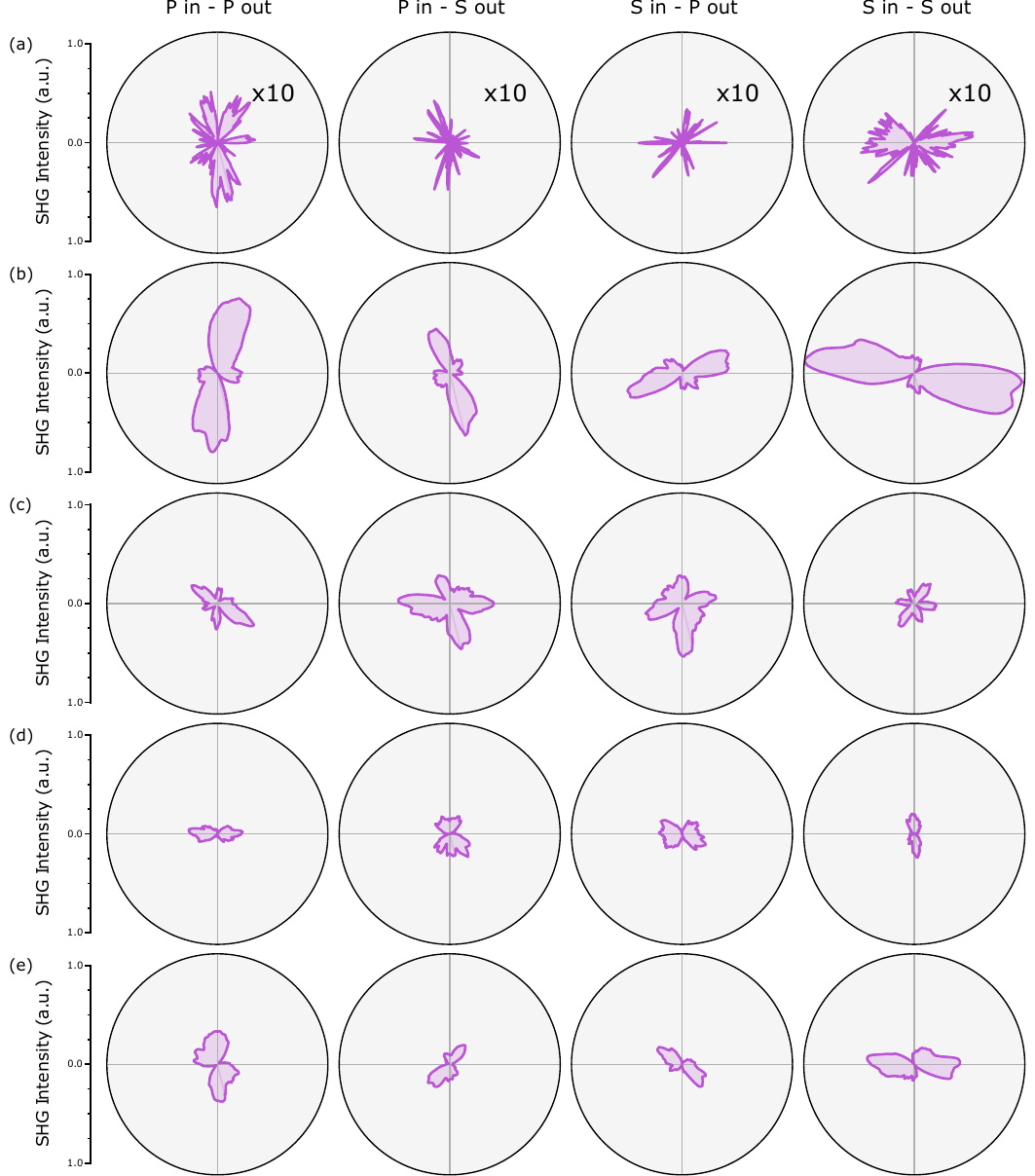}
\captionsetup{singlelinecheck=off}
\caption[]{
\label{fullequilibrium}
\begin{enumerate*}[label=\caplabel, ref=\capref]
\item {} \item {} \item {} \item {} \item \glsfmtshort{rashg} results of \cref{fig1:b,fig1:d,fig1:e,fig1:f,fig1:g} depicted in all four polarization channels (\PP, \PS, \SP, and \SS).
\end{enumerate*}
}
\end{figure}
 
\clearpage
\begin{figure}
\centering
\includegraphics{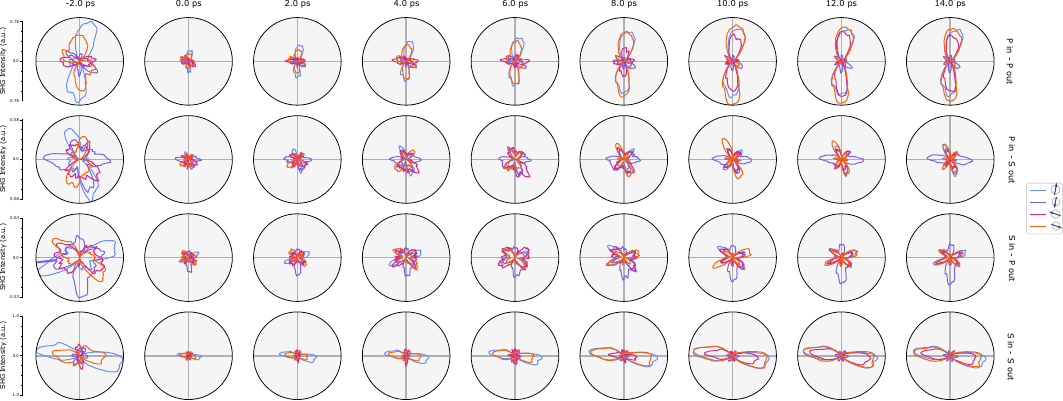}
\caption{\label{fullnonequilibrium}\glsfmtshort{rashg} results of \cref{fig2} depicted in all four polarization channels (\PP, \PS, \SP, and \SS).
The pump fluence is set to $\apx 600$ \si{\mu J \cdot cm^{-2}}.}
\end{figure}
 
\clearpage
\begin{figure}
\centering
\includegraphics{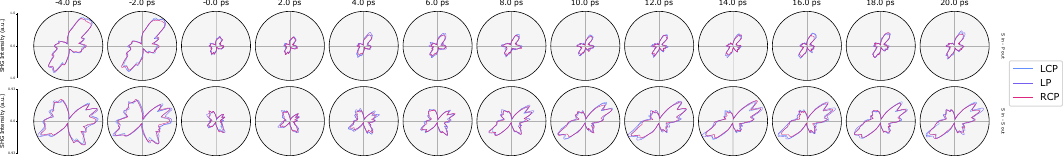}
\caption{\label{polarization}\glsfmtshort{rashg} results in a representative domain as a function of time for different pump polarization states (left circularly- (LCP), right circularly- (RCP), and linearly-polarized).
The pump fluence is set to $\apx 600$ \si{\mu J \cdot cm^{-2}}.
}
\end{figure}
 
\clearpage
\begin{figure}
\centering
\includegraphics{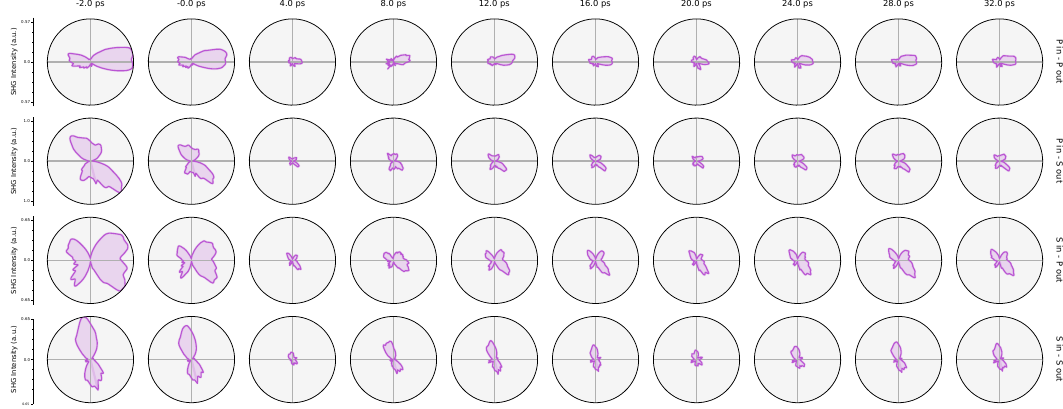}
\caption{\label{CtoBshort}\glsfmtshort{rashg} snapshots showing a transition to the state opposite to \figs{3}{c-d}.
The pump fluence is set to $\apx 600$ \si{\mu J \cdot cm^{-2}}.
}
\end{figure}
 
\clearpage
\begin{figure}
\centering
\includegraphics{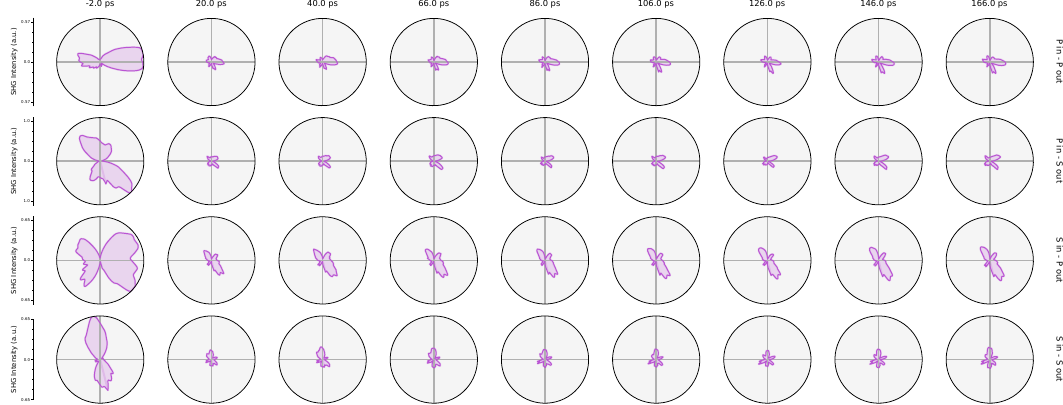}
\caption{\label{CtoBlong}\glsfmtshort{rashg} results in the same domain as \supsupcref{CtoBshort} plotted out to longer times.
The pump fluence is set to $\apx 600$ \si{\mu J \cdot cm^{-2}}.
}
\end{figure}
 
\clearpage
\begin{figure}
\centering
\includegraphics{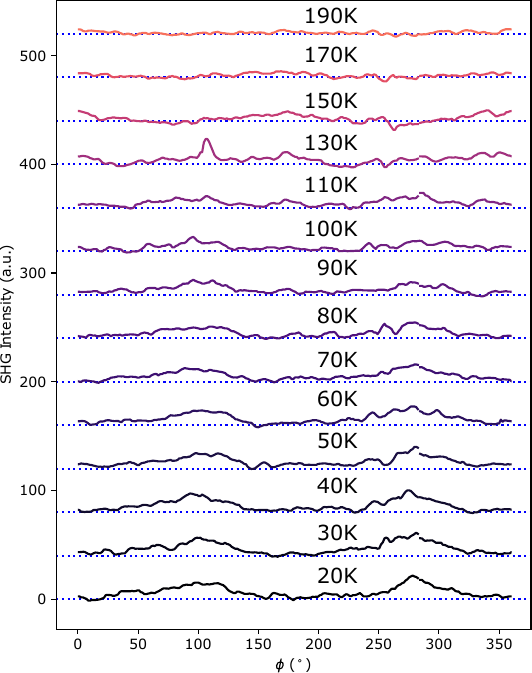}
\caption{\label{ctempdep}
SHG intensity in the same domain as \cref{fig1:f} as a function of temperature.
}
\end{figure}
 
\clearpage
\begin{figure}
\centering{
\phantomsubfloat{\label{domain_homogeneity:a}}
\phantomsubfloat{\label{domain_homogeneity:b}}
}
\includegraphics{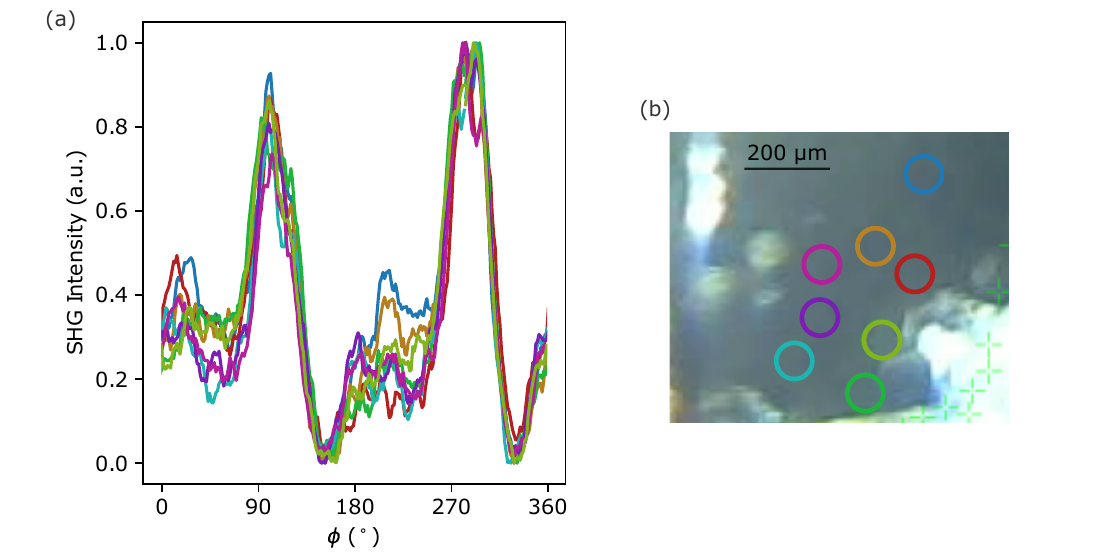}
\captionsetup{singlelinecheck=off}
\caption[]{
\label{domain_homogeneity}
\begin{enumerate*}[label=\caplabel, ref=\capref]
\item[] Demonstration of the homogeneity of the \gls{afm} domains in \cmb.
\item Normalized \Glsfmtshort{rashg} intensity in the \PP polarization channel in equilibrium at a variety of locations on the \cmb sample in \cref{fig1}.
\item Relative measurement locations (circles) depicted on a section of the \cmb sample imaged under white light illumination.
The size of the circles denotes roughly the beam spot size ($\sim 100$ \si{\mu m}), and the color of the circle borders refers to the color of the \gls{rashg} data in \ref{domain_homogeneity:a}.
\end{enumerate*}
}
\end{figure}
 
\clearpage
\begin{figure}
\centering{
}
\includegraphics{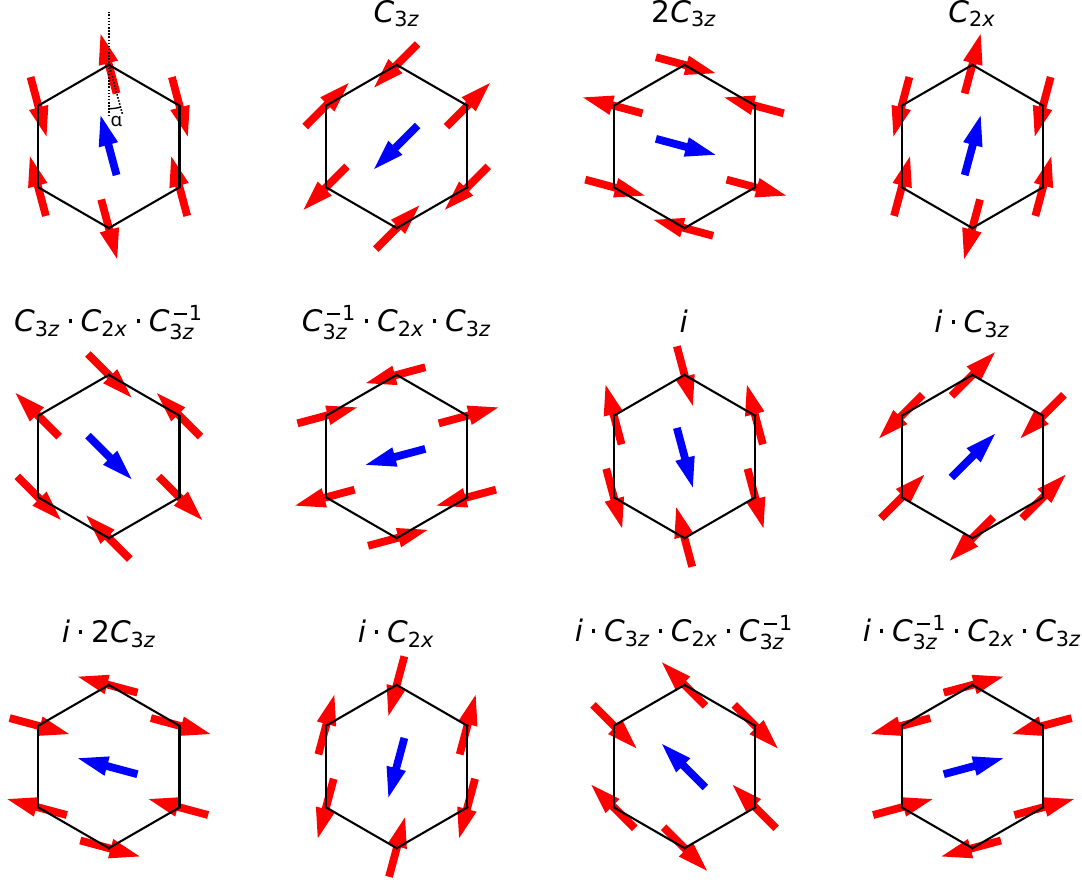}
\captionsetup{singlelinecheck=off}
\caption[]{
\label{fig:grouptheory}
Illustration of the equilibrium \ce{Mn} magnetic moment directions (red arrows) and \neel vectors (blue arrows, defined as the difference between two adjacent moments) in the twelve group theory-allowed domains in \cmb, generated by performing the labeled symmetry operation of the high temperature point group on the spin configuration shown in the top left corner.
Here, the $x$-axis is horizontal, and the $z$ axis is out of the page.
The angle $\alpha$ is known from neutron scattering\cite{gibson_magnetic_2015} to satisfy $0 < \alpha < \pi/6$, but is otherwise not quantiatively determined.
}
\end{figure}
 
\clearpage
\begin{figure}
\centering{
\phantomsubfloat{\label{transport:a}}
\phantomsubfloat{\label{transport:b}}
}
\includegraphics{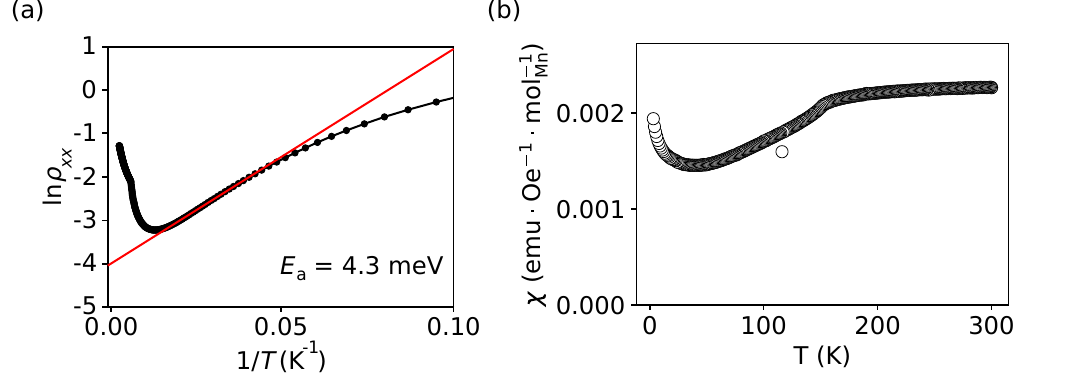}
\captionsetup{singlelinecheck=off}
\caption[]{
\label{transport}
\begin{enumerate*}[label=\caplabel, ref=\capref]
\item Log resistivity versus reciprocal temperature.
The red line is a linear fit showing activated behavior with $E_a=\qty{4.3}{meV}$.
\item Magnetic susceptibility versus temperature with $\mu_0 H=\qty{1}{T}~||~[011]$ showing the antiferromagnetic transition at $\sim\qty{150}{K}$.
\end{enumerate*}
}
\end{figure}
 
\clearpage
\begin{figure}
\centering
\includegraphics{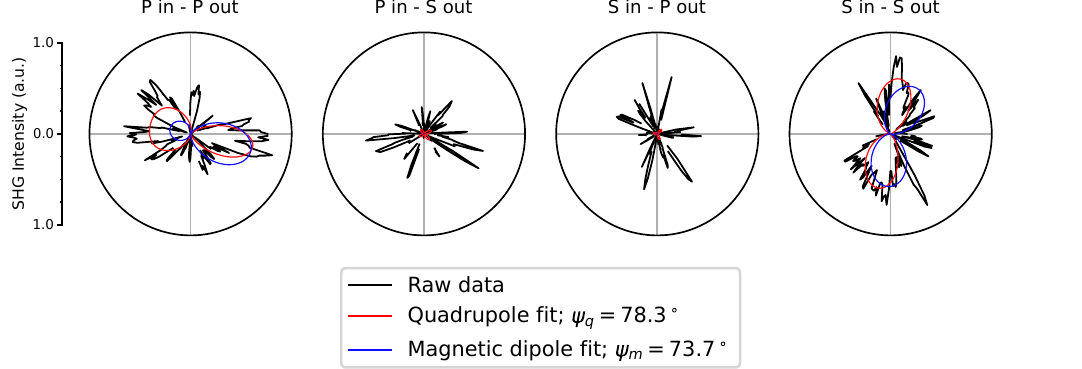}
\caption{\label{fig:orientation}High-temperature \glsfmtshort{rashg} data fit to quadrupole (red) and magnetic dipole (blue) models for orienting the sample $a$ axis with the $\phi=0$ plane (see \supcref{sec:orientation}).}
\end{figure}
 
\clearpage
\begin{figure}
\centering{
\phantomsubfloat{\label{interference:a}}
\phantomsubfloat{\label{interference:b}}
}
\includegraphics{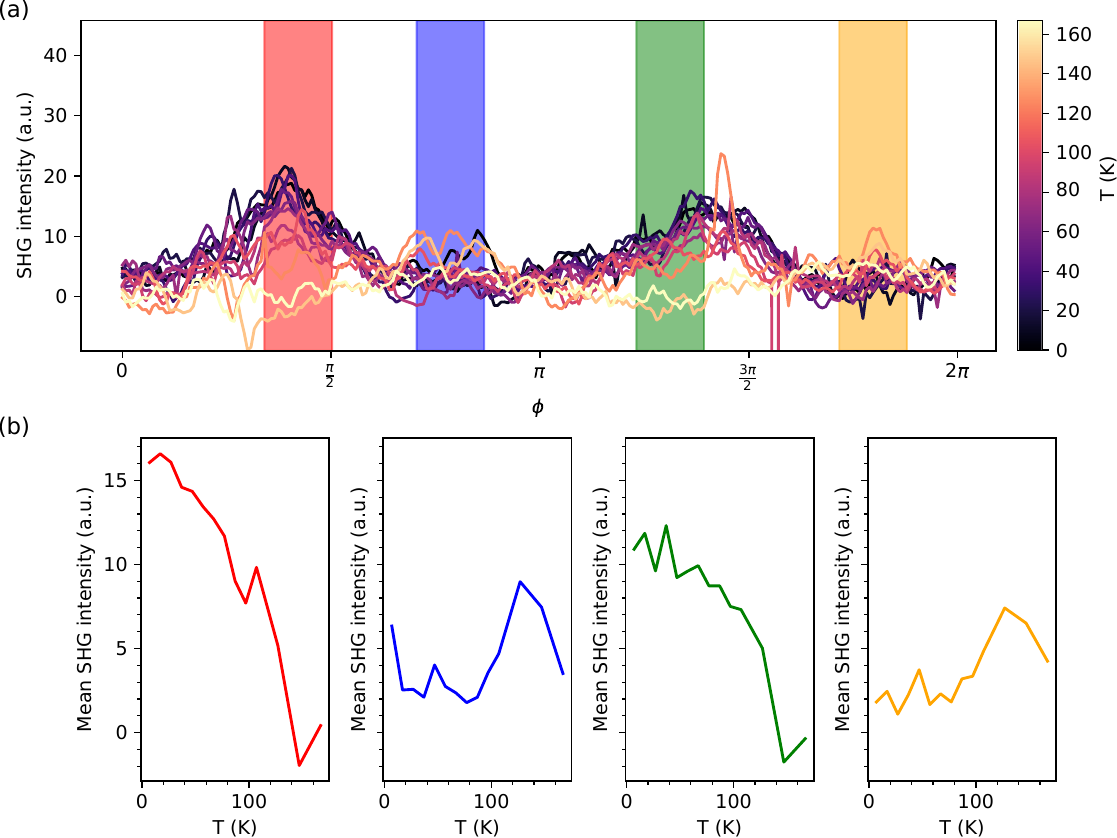}
\captionsetup{singlelinecheck=off}
\caption[]{
\label{interference}
\begin{enumerate*}[label=\caplabel, ref=\capref]
\item \Glsfmtshort{shg} intensity in the same domain as \cref{fig1:f} in the \SS polarization channel as a function of $\phi$ at various temperatures above and below $T_c$.
\item Means of the four shaded regions in \ref{interference:a} as a function of temperature, showing that the \glsfmtshort{shg} intensity for some values of $\phi$ decreases quickly below $T_c$ (to a minimum of $\sim 0$).
\end{enumerate*}
}
\end{figure}
 
\clearpage
\begin{figure}
\centering{
\phantomsubfloat{\label{thermalcycles:a}}
\phantomsubfloat{\label{thermalcycles:b}}
\phantomsubfloat{\label{thermalcycles:c}}
}
\includegraphics{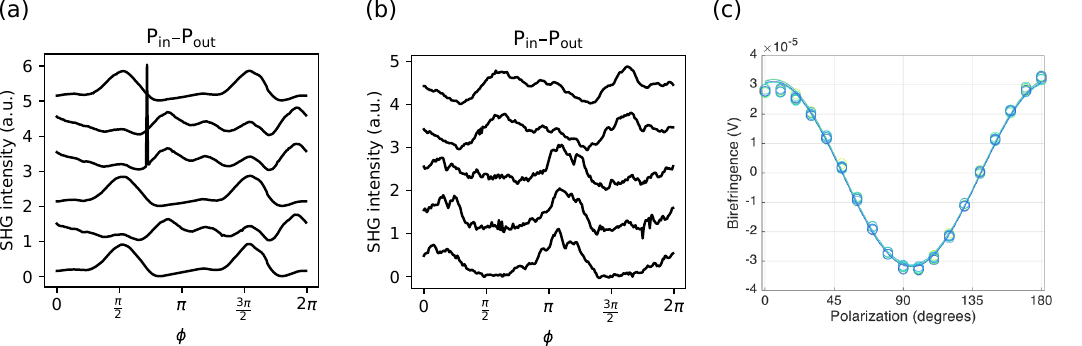}
\captionsetup{singlelinecheck=off}
\caption[]{
\label{thermalcycles}
\begin{enumerate*}[label=\caplabel, ref=\capref]
\item,\item \PP \Glsfmtshort{rashg} patterns in the ``left'' (a) and ``right'' (b) locations of the sample in \cref{fig1:c}, where each line corresponds to a different domain configuration obtained by cycling the sample temperature above and below $T_c$.
Data is normalized to \qty{1.0} and offset along the vertical axis for clarify.
\item \Glsfmtshort{ptmb} polarization plots at a single location for ten different thermal cycles above and below $T_c$, showing no change in the \glsfmtshort{ptmb} response upon thermal cycling.
The data in (a) and (b) were taken on the same sample as in \cref{fig1}, and the data in (c) was taken on second sample from the same batch.
\end{enumerate*}
}
\end{figure}
 
\clearpage
\begin{figure}
\centering{
\phantomsubfloat{\label{birefringence:a}}
\phantomsubfloat{\label{birefringence:b}}
\phantomsubfloat{\label{birefringence:c}}
\phantomsubfloat{\label{birefringence:d}}
\phantomsubfloat{\label{birefringence:e}}
\phantomsubfloat{\label{birefringence:f}}
\phantomsubfloat{\label{birefringence:g}}
}
\includegraphics{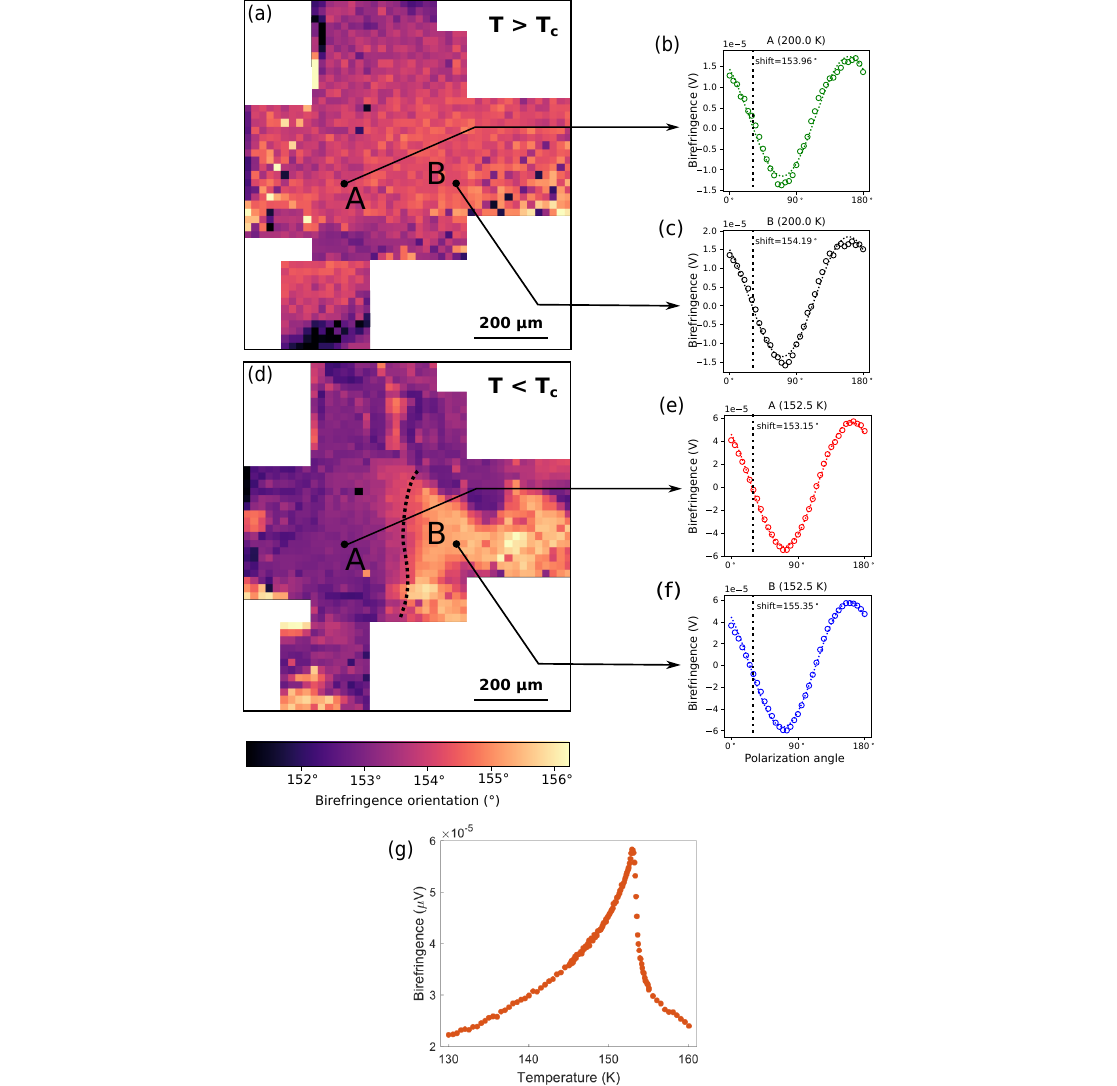}
\captionsetup{singlelinecheck=off}
\caption[]{
\label{birefringence}
\begin{enumerate*}[label=\caplabel, ref=\capref]
\item PTMB map above the transition temperature (duplicate of \cref{fig1:a}) with spots ``A'' and ``B'' labeled.
\item Polarization dependence of the birefringence signal in spot A at \qty{200}{K}.
\item Polarization dependence of the birefringence signal in spot B at \qty{200}{K}.
\item PTMB map below the transition temperature (duplicate of \cref{fig1:c}) with spots ``A'' and ``B'' labeled.
\item Polarization dependence of the birefringence signal in spot A at \qty{152.5}{K}.
\item Polarization dependence of the birefringence signal in spot B at \qty{152.5}{K}.
Vertical dotted lines in (b-c), (e-f) indicate the location of the crystallographic $a$-axis.
Dotted curves in (b-c), (e-f) are fits of the data to a single sine model.
The best fit phase shift parameter of that model is noted in the figure.
\item Temperature dependence of the birefringence signal at a single polarization angle and a single spot on the sample.
\end{enumerate*}
}
\end{figure}
 
\clearpage
\begin{figure}
\centering
\includegraphics{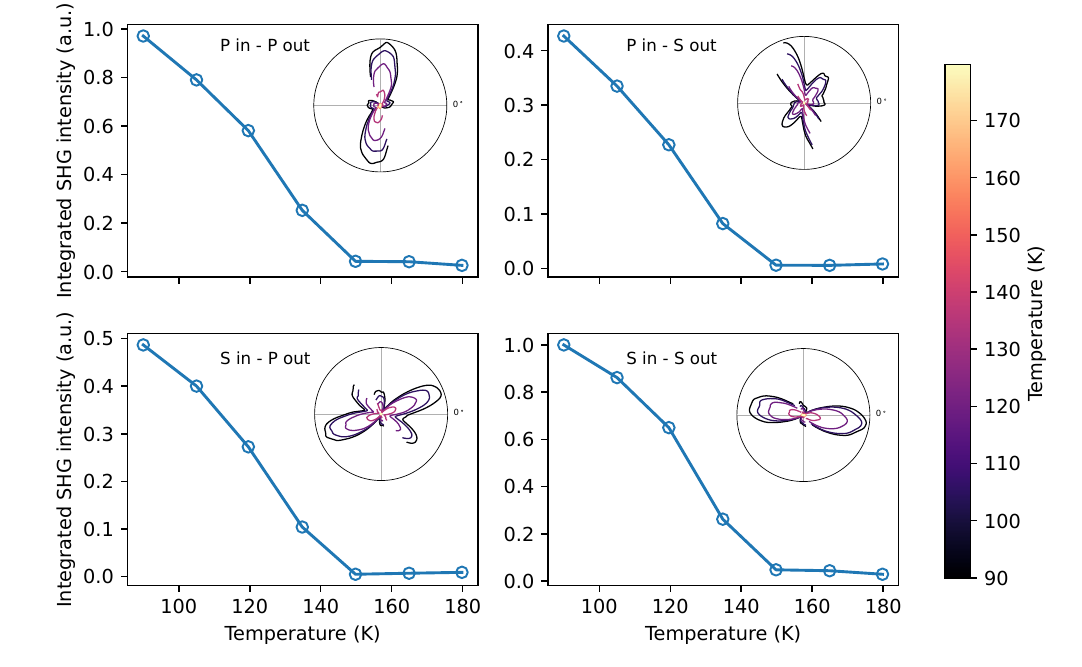}
\captionsetup{singlelinecheck=off}
\caption[]{
\label{pump-on-tempdep}
\begin{enumerate*}[label=\caplabel, ref=\capref]
\item[] Integrated SHG intensity for a representative \glsfmtshort{afm} domain in four polarization combinations (\PP, \PS, \SP, and \SS) as a function temperature, taken while the pump is on the sample.
Insets show the \glsfmtshort{rashg} patterns which are integrated at each temperature.
The pump fluence was \qty{300}{\mu J\cdot cm^{-2}}.
The difference in transition temperature compared to the un-pumped case (see \cref{fig0:b} inset) is negligible, suggesting that the static static heating due to the pump is small at the fluences discussed in this work.
\end{enumerate*}
}
\end{figure}
 
\end{widetext}

\end{document}